\documentclass[aps,prx,twocolumn,floatfix,superscriptaddress,longbibliography,10pt,draft]{revtex4-2}

\usepackage[utf8]{inputenc}
\usepackage[dvips]{graphicx}
\usepackage{amsmath,amssymb,amsthm,mathrsfs,amsfonts,dsfont,mathtools,physics}
\usepackage{epsfig}
\usepackage{braket}
\usepackage[colorlinks=true,citecolor=blue,linkcolor=magenta,final]{hyperref}
\usepackage[capitalize]{cleveref}
\usepackage{bm}
\usepackage{enumerate}
\usepackage{color}
\usepackage{graphicx}
\usepackage{tikz}
\usetikzlibrary{quantikz2,calc,shadings}
\usepackage{appendix}
\usepackage{nicematrix}
\usepackage{booktabs}      
\usepackage{bbm}

\usepackage{enumitem}
\usepackage[normalem]{ulem}
\usepackage{comment}
\usepackage{xcolor}

\definecolor{LB}{RGB}{134,41,198}%fiorentina color

\usepackage[framemethod=TikZ]{mdframed}
\definecolor{thmgray}{gray}{0.95}
\definecolor{conclusionbackground}{gray}{0.95}
\newcommand{\thmspaceafter}{2mm}

\newtheorem{myresult}{Myth}
\newenvironment{myth}%
{\begin{mdframed}[backgroundcolor=thmgray,nobreak,roundcorner=5pt,linewidth=1pt]\begin{myresult}}%
		{\end{myresult} \vspace{\thmspaceafter} \end{mdframed}}

\newtheorem{myconclusion}{Verdict}
\newenvironment{conclusion}%
{\begin{mdframed}[backgroundcolor=conclusionbackground,nobreak,roundcorner=5pt,linewidth=1pt]\begin{myconclusion}}%
		{\end{myconclusion} \vspace{\thmspaceafter} \end{mdframed}}

\usepackage{xcolor}

\definecolor{darkred}{RGB}{179, 57, 61}
\definecolor{darkgreen}{RGB}{83, 176, 142}

\newcommand{\gauge}[2]{
	\begin{tikzpicture}[scale=0.4]
		\def\startangle{180}
		\def\endangle{0}
		\def\radius{4}
		\def\innerradius{2.5}
		
		\draw[line width=2pt] 
		(\startangle:\radius) arc (\startangle:\endangle:\radius) --
		(\endangle:\innerradius) arc (\endangle:\startangle:\innerradius) -- cycle;
		
		\begin{scope}
			\clip (\startangle:\innerradius) arc (\startangle:\endangle:\innerradius) --
			(\endangle:\radius) arc (\endangle:\startangle:\radius) -- cycle;
			
            \fill[left color=darkred, right color=darkgreen]  
			(-4.,0) rectangle (4.,4.5);
		\end{scope}
		
		\draw[line width=1pt, black, rotate=-#1, fill=black] (0.15,0) -- (0,3.5) -- (-0.15,0) -- cycle;
		\fill[black] (0,0) circle (0.3);
			
		\node[font=\normalsize] at (-5.7,0) {Correct};
		\node[font=\normalsize] at (0,4.6) {#2};
		\node[font=\normalsize] at (6.1,0) {Debunked};
	\end{tikzpicture}
}

\usepackage{ifdraft}
\ifdraft{
	\newcommand{\authnote}[3]{{\color{#3} {\bf  #1} #2}}	
}{
	\newcommand{\authnote}[3]{}
}

\def\removefirstdot#1{\if.#1{}\else#1\fi}
\newcommand{\skp}[3]{#2}
\newcommand{\lName}{1}
\newcommand{\natcomm       }{\if\lName1\skp{  }{Nature Communications}{                              }\else{Nat. Comm.}\fi}
\newcommand{\siamjc        }{\if\lName1\skp{  }{SIAM Journal on Computing}{                       }\else{SIAM J. Comp.}\fi}
\newcommand{\quantum       }{\if\lName1\skp{  }{{Quantum}}{                                              }\else{Quant.}\fi}
\newcommand{\soda       }[1]{\if\lName1\skp{  }{Proceedings of the #1 {ACM-SIAM} Symposium on Discrete Algorithms ({SODA})}{                                    }\else{SODA}\fi}
\newcommand{\stoc       }[1]{\if\lName1\skp{  }{Proceedings of the #1 {ACM} Symposium on the Theory of Computing ({STOC})}{                                     }\else{STOC}\fi}

\begin{document}

\title{\texorpdfstring{Myths around quantum computation before
full fault tolerance:\linebreak
 What no-go theorems rule out and what they don't}{Myths around quantum computation before
full fault tolerance: What no-go theorems rule out and what they don't}}
\author{Zoltán Zimborás}\email{zoltan@algorithmiq.fi}
\affiliation{Algorithmiq Ltd, Kanavakatu 3C 00160 Helsinki, Finland}
\affiliation{HUN-REN Wigner Research Centre for Physics, 1525 P.O. Box 49, Hungary}
\author{B\'{a}lint Koczor}\email{koczor@maths.ox.ac.uk}
\affiliation{Mathematical Institute, University of Oxford, Woodstock Road, Oxford OX2 6GG, United Kingdom}
\affiliation{Quantum Motion, 9 Sterling Way, London N7 9HJ, United Kingdom}
\author{Zo\"{e} Holmes}\email{zoe.holmes@epfl.ch}
\affiliation{Institute of Physics, Ecole Polytechnique F\'{e}d\'{e}rale de Lausanne (EPFL), CH-1015 Lausanne, Switzerland}
\author{Elsi-Mari Borrelli}
\affiliation{Algorithmiq Ltd, Kanavakatu 3C 00160 Helsinki, Finland}
\author{András Gilyén}
\affiliation{HUN-REN Alfr\'{e}d R\'{e}nyi Institute of Mathematics, Budapest, Hungary}
\author{Hsin-Yuan Huang}
\affiliation{Google Quantum AI, Venice, CA 90291, USA}
\affiliation{IQIM, California Institute of Technology, Pasadena, CA 91125, USA}
\author{Zhenyu Cai}
\affiliation{Quantum Motion, 9 Sterling Way, London N7 9HJ, United Kingdom}
\affiliation{Department of Materials, University of Oxford, Parks Road, Oxford OX1 3PH, United Kingdom}
\author{Antonio Ac\'{i}n}
\affiliation{ICFO-Institut de Ciencies Fotoniques, The Barcelona Institute of Science and Technology, 08860 Castelldefels (Barcelona), Spain}
\affiliation{ICREA-Institucio Catalana de Recerca i Estudis Avan\c{c}ats, Lluis Companys 23, 08010 Barcelona, Spain}
\author{Leandro Aolita}
\affiliation{Quantum Research Centre, Technology Innovation Institute, Abu Dhabi, UAE}
\author{Leonardo Banchi}
\affiliation{Dipartimento di Fisica e Astronomia, Università di Firenze, Via Sansone, 1, 50019, Sesto Fiorentino (FI), Italy}
\affiliation{INFN Sezione di Firenze, Via Sansone, 1, 50019, Sesto Fiorentino (FI), Italy}
\author{Fernando G. S. L. Brand{\~a}o}
\affiliation{AWS Center for Quantum Computing, Pasadena, CA 91125, USA}
\affiliation{IQIM, California Institute of Technology, Pasadena, CA 91125, USA}
\author{Daniel Cavalcanti}
\affiliation{Algorithmiq Ltd, Kanavakatu 3C 00160 Helsinki, Finland}
\author{Toby Cubitt}
\affiliation{Phasecraft Ltd, London, UK}
\affiliation{Department of Computer Science, University College London, London, UK}
\author{Sergey N. Filippov}
\affiliation{Algorithmiq Ltd, Kanavakatu 3C 00160 Helsinki, Finland}
\author{Guillermo Garc\'{i}a-P\'{e}rez}
\affiliation{Algorithmiq Ltd, Kanavakatu 3C 00160 Helsinki, Finland}
\author{John Goold}
\affiliation{Algorithmiq Ltd, Kanavakatu 3C 00160 Helsinki, Finland}
\affiliation{School of Physics, Trinity College Dublin, Dublin 2, Ireland}
\affiliation{Trinity Quantum Alliance, Unit 16, Trinity Technology and Enterprise Centre, Dublin 2, Ireland}
\author{Orsolya Kálmán}
\affiliation{HUN-REN Wigner Research Centre for Physics, 1525 P.O. Box 49, Hungary}
\author{Elica Kyoseva}
\affiliation{NVIDIA Corporation, 2788 San Tomas Expressway, Santa Clara, 95051, CA, USA}
\author{Matteo A.C. Rossi}
\affiliation{Algorithmiq Ltd, Kanavakatu 3C 00160 Helsinki, Finland}
\affiliation{QTF Centre of Excellence, Department of Physics,
University of Helsinki, P.O. Box 43, FI-00014 Helsinki, Finland}
\author{Boris Sokolov}
\affiliation{Algorithmiq Ltd, Kanavakatu 3C 00160 Helsinki, Finland}
\author{Ivano Tavernelli}
\affiliation{IBM Quantum, IBM Research Europe -- Zurich, 8803 R\"{u}schlikon, Switzerland}
\author{Sabrina Maniscalco}
\affiliation{Algorithmiq Ltd, Kanavakatu 3C 00160 Helsinki, Finland}
\affiliation{QTF Centre of Excellence, Department of Physics,
University of Helsinki, P.O. Box 43, FI-00014 Helsinki, Finland}

\begin{abstract}
In this perspective article, we revisit and critically evaluate prevailing viewpoints on the capabilities and limitations of near-term quantum computing and its potential transition toward fully fault-tolerant quantum computing. We examine theoretical no-go results and their implications, addressing misconceptions about the practicality of quantum error mitigation techniques and variational quantum algorithms. By emphasizing the nuances of error scaling, circuit depth, and algorithmic feasibility, we highlight viable near-term applications and synergies between error mitigation and early fault-tolerant architectures. Our discussion explores strategies for addressing current challenges, such as barren plateaus in variational circuits and the integration of quantum error mitigation and quantum error correction techniques. We aim to underscore the importance of continued innovation in hardware and algorithmic design to bridge the gap between theoretical potential and practical utility, paving the way for meaningful quantum advantage in the era of late noisy intermediate scale and early fault-tolerant quantum devices.
\end{abstract}

\maketitle
\phantom{a}
\newpage
\phantom{a}
\newpage

\section{Introduction}

It has been six years since Preskill's visionary perspective article~\cite{preskill2018QuantCompNISQEra} introduced the concept of Noisy Intermediate-Scale Quantum (NISQ) technology, marking the beginning of a new era in quantum computation. The initial excitement surrounding near-term quantum computation led to unrealistic expectations. As the field matured, this enthusiasm has been tempered by extensive theoretical research proving significant limitations on NISQ-friendly methods \cite{takagi2022fundamental,deshpande2022tight, takagi2023universal, tsubouchi2023universal, quek2022exponentially}. These results have influenced the community's perspective, contributing to an increasingly negative connotation around such techniques.

Yet, deconstructing unrealistic expectations should not lead to a dismissal of real opportunities. It is crucial to recognize the scope of theoretical no-go results, which often pertain only to the asymptotic regime and have sometimes been misinterpreted as definitive statements about practical performance. The purpose of this perspective is to reevaluate how we think about quantum computing before full fault tolerance is achieved and to reconcile the two strikingly different pictures that theoretical proofs and optimistic practical aspirations paint.

While the current stage of quantum technology is often described as the NISQ era, many researchers argue that we are still in its early phase; and it may take until what some call the 'late NISQ era' to witness quantum computations achieving practical significance. 
This article's key purpose is to demonstrate how quantum hardware will likely evolve through multiple natural stages, forming a relatively continuous transition from the early to the late NISQ era, then to early fault tolerance, and eventually to full fault tolerance.

In this perspective, we aim to debunk common beliefs, or ``myths,'' surrounding techniques such as error mitigation and variational quantum algorithms. We explain how tools initially developed for non-error-corrected devices can naturally extend to early error-corrected machines, enabling a gradual and synergistic integration into the broader landscape of quantum computing. We also identify problems where these methods may lead to viable practical applications in the near term. Inevitably, any concrete statement of these myths invites criticism that we may be oversimplifying very complex questions or exaggerating certain views. When precisely stating those views, our intention is merely to put forward these perspectives as focal points to clarify the debate.

\section{Myths}

\subsection{Utility of Quantum Error Mitigation}
Quantum error mitigation (QEM), in contrast to quantum error correction (QEC), does not rely on the concept of a logical qubit, where information is encoded across many physical qubits and protected against errors through complex correction protocols. Instead, QEM \cite{temme2017error, endo2018practical, Kim_scalable_2023,filippov2023scalable,PhysRevX.11.031057, PhysRevX.11.041036, czarnik2021error, maciejewski2020mitigation, van2022model, cai2023quantum} operates directly on physical qubits, performing computations without the need for extensive encoding.
It builds on the observation that most near-term algorithms estimate
expectation values by repeatedly measuring the output of a quantum circuit, and generally aim to reduce the bias that is introduced by noise.
However, the resulting rectification process trades off bias for increased variance in the estimator, which consequently increases the number of samples required to achieve accurate results. 
In this paper, we will mostly focus on qubit-based quantum computers, although similar arguments apply
to other platforms, e.g., based on photons \cite{su2021error,taylor2024quantum,faurby2024purifying}.
It has been shown that the sampling overhead for any error mitigation method grows exponentially with the size of the quantum circuit. In this subsection, we consider two commonly quoted views regarding the limitations of error mitigation that
are based on the theoretically shown scaling limits and argue why these claims may not be justified.

\subsubsection{Exponential scaling of error mitigation methods}
\begin{myth}
Quantum error mitigation cannot be useful due to the scalability limitations revealed by recent no-go theorems.
\end{myth}

Imagine a very simple noise model, whereby a quantum circuit consists of $G$ gates and each gate has a probability $\varepsilon$ that its execution is faulty. Then, using elementary probability theory, the probability that \textbf{all} gates are executed perfectly is exactly $(1{-}\varepsilon)^G \sim e^{- \varepsilon G}$.
This simple argument already suggests an exponential decay of quantum information in terms of the number of gates $G$.
Although error mitigation techniques were initially met with some over-enthusiasm, the simple argument outlined above led to early recognition that exponential costs are to be expected. This anticipation of scalability challenges in quantum error mitigation was already present in its early days, and subsequent no-go theorems reinforced these initial expectations. In the following, we briefly review some of the rigorous theoretical results that imply that indeed quantum information degrades exponentially as the depth $L$ and width $N$ ($NL/2 =G$) of the underlying circuits increase. 

\begin{table*}[tb]
    \centering
    \begin{tabular}{|l|c|c|c|c|c|c|}
\hline
      &  Google Willow \cite{Acharya_2024_N} & IBM Heron r2 \cite{IBM_QPUs} & Quera \cite{evered2023high} & IonQ Forte \cite{chen2023benchmarking}& Quantinuum H2 \cite{H2}&
      Oxford Ionics \cite{loschnauer2024scalable}\\
\hline 
       $\varepsilon$ &  $3.3 \times 10^{-3}$    &  
$2.2 \times 10^{-3}$  & $5 \times 10^{-3}$  &  $4 \times 10^{-3}$  &  $8.5 \times 10^{-4}$ & $2.9 \times 10^{-4}$ \\
\hline 
     $N$ &  105 & 156 &  256 & 36 & 30 & 10\\
\hline
    \end{tabular}
    \caption{Number of qubits  $N$  and average two-qubit gate errors $\varepsilon$ for  current quantum devices. }   \label{fig:gate_errors}
\end{table*}

The simplest rigorous argument
to show that error mitigation would require an exponential number of samples
is based on the observation that under certain noise models the output state rapidly converges to the maximally mixed state \cite{muller2016relative, stilck2021limitations, deshpande2022tight}, implying that just to distinguish the resulting state from the maximally mixed state one would already need exponentially many samples.
More sophisticated information-theoretic tools were employed to analyse the sample-overhead under more general noise models in Refs.~\cite{takagi2022fundamental,deshpande2022tight, takagi2023universal, tsubouchi2023universal, quek2022exponentially}.
The common approach in their theoretical analysis is the following: When one applies a noisy circuit to two arbitrary input states, the outputs become drastically less distinguishable from each other.
Thus, to obtain any meaningful result, e.g., through mitigation methods, the required sample overhead has to scale exponentially with the total number of gates. 

For example, Ref.~\cite{tsubouchi2023universal} provides an $\Omega(\exp(\varepsilon N L))$ lower bound on the required sample overhead for brickwork circuits with $N$ qubits and depth $L$ featuring layer-wise local depolarizing noise. This fundamental lower bound matches well the known sample overheads of 
state-of-the-art error mitigation techniques \cite{temme2017error,Kim_scalable_2023,filippov2023scalable,PhysRevX.11.031057, PhysRevX.11.041036}, each associated with a sampling cost of $p(\varepsilon  N L)\,$exp$(\varepsilon B  N L)$, where $p$ is some polynomial function and $B$ is some constant; see the particular values in Refs. \cite{cai2023quantum,filippov2024scalability}. For example, both the Zero Noise Extrapolation and Tensor Network Error Mitigation methods yield for the exponential prefactor $B=1$.

We argue that the multiplicative factor \textbf{$\varepsilon$ in the exponent may be very small and could allow for a large number of gates $ NL \propto \varepsilon^{-1} $} and thus, despite the exponential scaling, there can be a finite regime, where quantum error mitigation may be feasible.
Thus, the fact that error mitigation methods scale exponentially does not mean that they are, per definition, useless.
The usefulness depends on the multiplicative factor $\varepsilon$ that represents gate error rates. In \cref{fig:gate_errors} we show gate error rates which have already been achieved --  these are expected to continue to improve as technological and benchmarking techniques \cite{Proctor_2024} evolve. Such improvements are necessary not because of stringent requirements on NISQ, but also because for efficient QEC one needs lower error rates.
Furthermore, as early error corrected machines become available, as we will explain later, error mitigation techniques remain relevant, and $\varepsilon$ will be a logical error rate that will be further reduced from current physical error rates as code distances are increased.
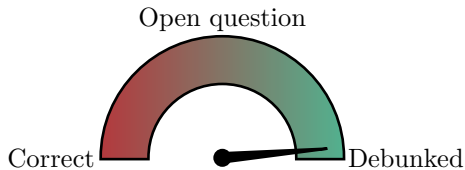
\begin{center}
    \gauge{85}{Open question}
\end{center}
\begin{conclusion}
    Error mitigation techniques scale exponentially with system size; however, the gate error rate $\varepsilon$ appears as a prefactor in the exponent, making it feasible for $\mathcal{O}(\frac{1}{\varepsilon})$ circuit sizes.
\end{conclusion}

\newpage

\subsubsection{Feasible circuit sizes in near-term quantum computing}
\begin{myth}
Solving practical problems requires much larger circuits than is feasible on quantum hardware without error correction capabilities.
\end{myth}

In the previous section, we concluded that in the presence of $\varepsilon$-noise, circuits of size $\mathcal{O}(\varepsilon)$ can be executed without prohibitive overhead. In fact, circuit sizes allowed by current error rates already led to non-trivial computations, as evidenced by experiments that could not be simulated using conventional computers, at least using methods available at the time they were conducted \cite{arute2019quantum,Zhong_2020, Morvan_2024}. However, these typically involve random sampling tasks that do not have an obvious practical application.
So, the question naturally arises: \textbf{Can these small circuit sizes yield any useful function in practical applications?}

We now estimate feasible circuit sizes informed by error mitigation formulas in order to identify potential applications. Currently we are entering an era where two-qubit (2Q) gate errors are already of the order $10^{-3}$ (see Table 2) and experimental progress steadily moves this towards $10^{-4}$. For superconducting devices, errors on the order of $10^{-5}$ may be possible, but two-qubit gate errors beyond that would be unrealistic with current technology.

Typically, with superconducting qubit devices, it is possible to run up to 1000-5000 circuit layer operations per second \cite{gambetta2021driving}. This means that even for a circuit depth of 100-1000, it is feasible to measure $10^{6}-10^{7}$ samples in a reasonable time. (Note that for ion traps and neutral atom devices the quantum gate execution times are orders of magnitude slower, resulting in lower feasible sample sizes.)
To get some crude estimates for feasible circuit sizes for algorithms using such sample sizes, we declare the affordable sampling overhead as $\mathrm{e}^{\varepsilon N L} \sim \mathrm{e}^{10}$
(note that $\mathrm{e}^{10} \approx 22026$), or in other words, the average number of errors in the circuit is of the order of $10$ and $N L \sim /\varepsilon$.  Considering the errors of the current and near-future devices, we obtain Table~\ref{circuit-sizes} about the reachable useful circuit sizes.

\newpage

\begin{table}[ht!]
    \centering
    \begin{NiceTabular}{|l|c|c|c|}
\midrule
     Average (2Q) error  & $\varepsilon=10^{-3}$ & $\varepsilon=10^{-4}$ & $\varepsilon=10^{-5}$  \\
\midrule
        Feasible circuit sizes  & $100 \times 100$ & $300 \times 300$, &  $1000 \times 1000$, \\
         &  & $100 \times 1000$  &  $100 \times 10000$  \\ 
\midrule
    \end{NiceTabular}
    \caption{Reachable useful circuit sizes for devices with different average two-qubit errors (assuming sampling frequency of 1 kHZ).}
    \label{circuit-sizes}
\end{table}

On pre-fault-tolerant architectures, physical error rates set a firm limit on executable circuit sizes. For example, $100$ qubit circuits with depth beyond $10^5$ would require reaching error rates below $10^{-6}$, which is currently beyond the roadmap of any hardware provider. Therefore, the focus of any realistic effort towards the near-term quantum advantage should be on reducing the circuit sizes for practical application cases to fit those presented in Table~\ref{circuit-sizes}. We foresee that by minimizing circuit sizes and using optimal error mitigation strategies, it will still be possible to perform quantum dynamics and many-body experiments with a practical advantage in the near term, but innovations in reducing circuit complexities for such applications play a crucial role in achieving this~\cite{Smith2019,Clinton2021,Clinton2024}.

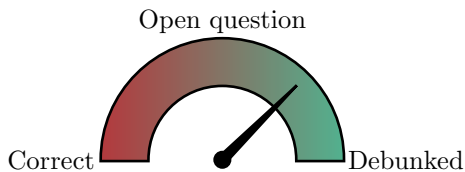
\begin{center}
    \gauge{45}{Open question}
\end{center}
\begin{conclusion}
 Pre-fault-tolerant circuit sizes may enable useful quantum applications.
 However, practical quantum advantage remains to be demonstrated.
\end{conclusion}

\subsubsection{Path towards fault tolerance\label{sec:path_towards_fault_tol}}

\begin{myth}
Quantum error mitigation and quantum error correction are two different worlds. 
As we pass the breakeven point, error correction will make error mitigation obsolete. 
\end{myth}

The threshold theorem is very powerful and guarantees that, beyond the breakeven point, scaling the size of an error-correcting code will enable exponentially low logical error rates. However, it is anticipated that the error rates will decrease rather slowly, since scaling up devices while maintaining physical error rates may prove to be a challenging experimental task.
Thus, in the first-generation error corrected devices, the fact that physical qubits are in short supply, while error rates are not significantly below the threshold, will mean that one must make a compromise to have sufficiently many logical qubits at the expense of a non-negligible logical error rate (e.g., $10^{-6}$). This is what we refer to as the early fault-tolerant quantum computing era. 

Although \textbf{existing QEM methods} were originally developed to tackle physical errors, they \textbf{can naturally be extended to target the non-negligible logical errors expected in the early fault-tolerant era} simply by replacing all the physical operations and errors by their logical counterparts. However, it should be noted that there are practical differences between tackling physical and logical errors.
In the following, we illustrate this on the example of PEC, which requires noise characterisation.
First, logical error rates are smaller and logical operations are slower, thus the logical noise characterization is harder in the general case. On the other hand, error models of certain logical components can be known beforehand, e.g., in magic state distillation~\cite{piveteauErrorMitigationUniversal2021},
significantly simplifying noise characterisation and recovery operations.
Moreover, recovery operations in logical PEC can be further simplified by using the Pauli basis which can be applied almost freely in a virtual way via Pauli frame updates~\cite{suzuki2022quantum}.

Another family of error mitigation techniques that has additional benefits when applied in the logical context is purification-based methods~\cite{PhysRevX.11.031057,PhysRevX.11.041036}, given that they do not require noise characterisation and the fact that logical errors are typically less coherent than physical errors. Error-corrected variants of zero-noise extrapolation have also been explored, e.g., extrapolation to infinite code distances~\cite{wahl2023zero}.

In addition to tackling errors due to device noise, quantum error mitigation can also be used to overcome compilation and algorithmic errors, some of which arise from constraints due to fault-tolerant implementations. These are errors that cannot be addressed using quantum error correction, even with perfect logical gates. For example, many quantum algorithms require continuous rotations; however,
precisely synthesising them is a challenge as they require potentially deep sequences of Clifford and T rotations, whereas techniques known from PEC allow one to synthesise rotations exactly on average~\cite{suzuki2022quantum,PhysRevLett.132.130602, koczor2024sparse,kliuchnikov2023shorter}. Furthermore, even in the absence of noise, one may need to face so-called algorithmic errors. For example, early fault-tolerant devices may enable intermediate-depth time evolutions using Trotter circuits, and extrapolation-based ideas can naturally be applied to suppress the effect of finite Trotter step size~\cite{PhysRevA.99.012334,rendon2024improved, chan2022algorithmic} or even to completely eliminate it~\cite{kiumi2024te}.

There are also ways to integrate QEC and QEM beyond concatenation. In Ref.~\cite{liu2024virtualchannelpurification}, the authors proposed a circuit to virtually entangle the noise in two different registers. If one encodes one of the registers in some QEC code and performs the parity checks, this can reveal the errors happening in the other unencoded register due to the noise entanglement, allowing one to correct the errors on the unencoded register. This circuit can also incorporate a bit-flip code and a phase-flip code to achieve the error correction power of a surface code, by merging the error correction power of the input codes.
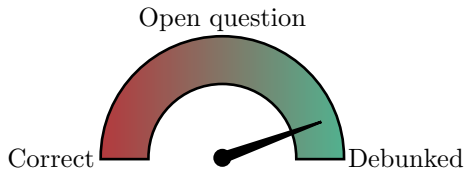
\begin{center}
    \gauge{70}{Open question}
\end{center}
\begin{conclusion}
Different generations of quantum hardware will likely undergo a continuous evolution and quantum error mitigation will be a major enabler in all evolutionary stages until full fault tolerance is reached.
\end{conclusion}

\subsection{Variational quantum algorithms}

Classical variational methods have long proven an effective method for solving computational problems both in physics and beyond~\cite{carleo2019machine}. It is thus entirely natural to ask whether the success of classical optimisation methods and modern developments in machine learning can be combined with the speedups expected of quantum algorithms. Motivated as such, variational quantum algorithms (VQAs)~\cite{cerezo2020variationalreview, endo2021hybrid} and quantum machine learning (QML)~\cite{biamonte2017quantum,schuld2018supervised,cerezo2020variationalreview, cerezo2022challenges,endo2021hybrid} have attracted an immense amount of attention over the past ten years. 

Variational quantum computing inherits many of the strengths and weaknesses of classical optimization-based algorithms. On the one hand, their flexibility both makes them applicable to broad range of problems and means they can more easily be adapted to the constraints of near-term hardware. On the other hand, they are fundamentally heuristic. In particular, and in contrast to more conventional quantum algorithms, these approaches generally lack guarantees on their resource requirements. This, combined with the lack of immediately available hardware to test their capabilities (in contrast to the classical case), makes it challenging to predict the viability of scaling them to interesting problem sizes. 

\subsubsection{Fundamental limitations of variational quantum algorithms}

\begin{myth}
    Variational quantum algorithms require exponential training efforts due to barren plateaus and thus cannot be useful.
\end{myth}

Over the past few years, a large community of researchers have attempted to analytically investigate whether or not variational quantum algorithms will be trainable on moderate-sized problems. A number of different barriers to trainability have been identified including poor local minima~\cite{bittel2021training,fontana2022nontrivial,anschuetz2022beyond,anschuetz2021critical,larocca2021diagnosing}, expressivity limitations~\cite{tikku2022circuit}, and the barren plateau phenomena~\cite{mcclean2018barren, larocca2024review}. Of these issues, perhaps in virtue of being more amenable to analytic studies, the overwhelming majority of attention so far has focused on barren plateaus.

A loss landscape is said to exhibit a barren plateau when its loss values become exponentially concentrated with problem size. In this case, loss gradients at any randomly chosen parameter value will with high probability be vanishingly small. Hence, for an overwhelming majority of parameter choices, exponentially many measurement shots are needed to identify the loss minimizing direction.

To put some rough numbers on the barrier posed by barren plateaus for different circuit sizes, we extrapolate the numerical variance scalings found in Refs~\cite{mcclean2018barren, letcher2023tight} for a 1D hardware efficient ansatz with a local loss and then apply Chebyshev's inequality to estimate an upper bound on the probability that any randomly chosen point has gradients greater than $\varepsilon = 10^{-3}$. For a $100\times50$ circuit we find that the probability is a potentially manageable $10^{-4}$ but for a $100\times1000$ circuit this probability drops to $10^{-24}$. We stress that these numbers are based on a crude numerical extrapolation (and thus potentially do not adequately account for the role of constant overheads); nonetheless, they paint a pessimistic picture of the prospects of training unstructured circuits with random initializations except at relatively modest circuit sizes.   

While there have been a number of approaches that have been proposed to avoid barren plateaus, it has recently been argued that in all those standard cases (that can be proven to avoid them) the resulting training process is in a sense classically simulable~\cite{cerezo2023does}. This can in fact be viewed as something positive that brings the prospects of a form of variational quantum computing nearer term. Namely, a quantum computer might be necessary in an initial data collection phase~\cite{bermejo2024quantum}, be required for inference ~\cite{khosrojerdi2024learning} or to eventually sample solutions (e.g., in the case of generative modeling), but the training could be performed classically. 
On the other hand, there could also exist the ``opposite'' learning scenario, where efficient training requires a quantum computer, while testing new data can be performed using purely classical means~\cite{jerbi2024shadows}.

Nonetheless, the barrier posed by barren plateaus currently attracts much debate. Of particular importance is the fact that barren plateaus are an average-case notion. In reality, practitioners do not care about most parameter values but rather the ones in the region around an initialization and along a trajectory to a reasonable problem solution, and such trajectories with good gradients could \textit{theoretically} exist on a barren plateau landscape. It remains an open question whether or not such trajectories can be found in practice.

In parallel, there are a number of physically-motivated Ans\"atze, that partially fall outside of current formal analyses, that may have favourable regimes for training at modest problem sizes. These include the chemically motivated unitary coupled cluster ansatz~\cite{peruzzo2014variational} and some variants to the Hamiltonian Variational Ansatz~\cite{wecker2015progress, hadfield2019quantum} (and the closely related Quantum Approximate Optimization Algorithm~\cite{farhi2014quantum}) which can be seen as digital relaxation of adiabatic ground state preparation protocols.

Another common obstacle to variational quantum circuit training is due to the complexity of the cost function landscape, which often shows many local minima where the optimizer can get stuck. Intuitively, this can be understood, since the expectation values of variational quantum circuits can be expanded to a Fourier series \cite{schuld2021effect}, with many oscillatory terms.

A classical solution to mitigate poor cost function landscapes, at least in machine learning applications, is to use overparametrized models where the number of trainable parameters is much larger than the number of data. Overparametrized models, when trained with gradient descent optimizers, offer several theoretical advantages: i) local minima often become saddle points and, overall, the gradient contains more ``directions'' to minimize the cost function; ii) for some models, it can be formally shown that \textit{most} local minima become \textit{good enough} \cite{anschuetz2021critical}, namely that their cost gets close to that of the global optimum.

There are several obstacles in trying to mimic the success of overparametrized models in the quantum setting. For a start, overparametrization may require either very deep or very wide (or both) quantum circuits \cite{larocca2023theory}, which might be challenging for near term devices and may lead to other other problems, e.g., barren plateaus.

Moreover, the remarkable success of classical overparametrized models is also due to simplicity of training 
them via the back-propagation algorithm, which reuses intermediate gradient calculations and makes the overall 
update step efficient. In the quantum setting, it is hard to reuse intermediate computational results due to no cloning. Nevertheless, back-propagation scaling has been obtained using a few copies of the quantum states \cite{abbas2024quantum} at the expense of poor precision scaling, which might be unavoidable~\cite{gilyen2017OptQOptAlgGrad}. Due to this extra ``copy complexity'', the degraded precision scaling, and the difficulty of squeezing many independent parameters into NISQ quantum circuits, it is currently an open problem to understand how to best reach overparametrization in NISQ-friendly scenarios.

Finally, to return to our starting point, the heuristic nature of variational quantum algorithms makes predicting their scalability challenging and it could turn out that, once better devices are available, these algorithms in practice work much better than can be guaranteed analytically. This is indeed the case for many classical optimization-based algorithms. In this manner, while the aim of this document is to assess the prospects of near-term quantum computing, variational quantum algorithms is one area where we may just have to wait and see. 

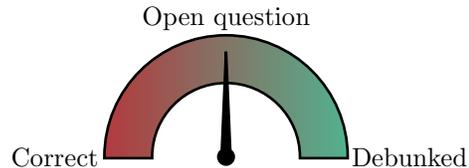
\begin{center}
    \gauge{0}{Open question}
\end{center}
\begin{conclusion}
While training unstructured quantum circuits at scale faces strong obstacles, the prospects of some problem-inspired models equipped with special initializations remain undetermined. Non-variational quantum subroutines could also potentially enhance classical variational methods. 
\end{conclusion}

\subsubsection{Variational quantum algorithms beyond NISQ}

\begin{myth}
    Variational quantum algorithms are only relevant in the NISQ era.
\end{myth}

Variational quantum algorithms were initially motivated, at least in part, by their comparative suitability to near-term hardware. In contrast to more conventional algorithms, it was anticipated that they could be run more flexibly with shorter circuits with more restricted topologies. This has led to the impression among some that variational methods are exclusively relevant in the NISQ era. 

However, as discussed above, independently of their (un)suitability to near-term hardware, it is entirely reasonable to ask whether or not variational components can be combined with quantum computing. After all, variational methods have a long proven track record in classical computing and are immensely popular in a broad range of practical applications, such as machine learning or quantum chemistry. In particular, non-variational methods (e.g., quantum phase estimation) often require a reasonable approximation to some target state (e.g., the ground state) to initiate the algorithm. In general, it is far from obvious how such approximations can be found and variational methods are a natural option to be used for this purpose. 

Nonetheless, there remain open questions on how to best implement variational optimization in fault tolerant machines but the field is gaining traction. One pressing issue is the fact that most variational approaches consider continuous optimization of rotation angles. The fine-grained resolution required here may in general lead to resource-intensive circuits with high T gate counts~\cite{sayginel2024faulttolerant}. Nevertheless, we refer to \cref{sec:path_towards_fault_tol} that quantum error mitigation techniques can be extended naturally to mitigate such circuit compilation errors (fine rotation angle resolution)~\cite{suzuki2022quantum,PhysRevLett.132.130602, koczor2024sparse,kliuchnikov2023shorter}
as well as algorithmic errors, such as finite Trotter approximation errors
~\cite{PhysRevA.99.012334,rendon2024improved, chan2022algorithmic},
and lead to significantly reduced T gate costs~\cite{kiumi2024te}. Furthermore, fault-tolerance may enable the use of amplitude amplification to estimate expected values with an improved $O(1/\varepsilon)$ scaling which is the central task in typical variational algorithms and usually the time bottleneck as standard estimation with quadratic dependence $O(1/\varepsilon^2)$ may require a large number of shots in practice~\cite{grinko2021iterative,PRXQuantum.5.010324}. On the other hand, fault tolerant gates will inevitably have slower clock rates than their physical counterparts and it becomes even more important to minimise the number of circuit repetitions whereby classical shadows may be an enabling tool~\cite{Huang_2020,PhysRevX.12.041022,PRXQuantum.5.010324}.
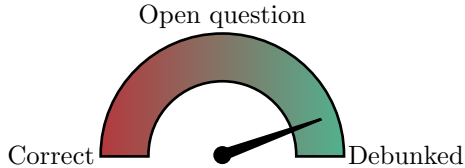
\begin{center}
    \gauge{70}{Open question}
\end{center}
\begin{conclusion}
	$\!$While some technical challenges,
such~as high circuit repetition counts and fine rotation-angle resolution,
need more attention, the community is making progress in addressing these, and some form of variational quantum algorithms will likely find useful applications in the fault-tolerant quantum computing era;
much like in classical computing wherevariational methods are very prominent.
\end{conclusion}

\subsection{Exponential speedups in applications}

\begin{myth}\label{mythExp}
We do not yet have \textbf{proven exponential} quantum speedups for end-to-end applications in machine learning, optimization, quantum chemistry, or materials science that guarantee substantial commercial and financial value.
\end{myth}

While quantum algorithms like Shor's factoring algorithm provide strong evidence for superpolynomial speedups in computer algebra problems \cite{shor1994Factoring, kitaev1995QMeasAndAbelianStabilizer}, commercial applications, which typically lack such clean mathematical structure, present a fundamentally different situation \cite{dalzell2023QuantumAlgSurvey,lee2023EvidenceQuantAdvGroundState}.

Quantum chemistry offers an illuminating example: Although simulating quantum dynamics is BQP-complete and thus is widely believed to be intractable for classical computers, real molecules have constant sizes and specific chemical properties. Molecular systems studied today, such as FeMoCo (the iron-molybdenum cofactor in nitrogenase), appear intractable for current classical methods, yet are expected to be tractable on modest-sized  quantum computers \cite{dalzell2023QuantumAlgSurvey}. For such constant-sized problems of practical interest, the traditional framework of asymptotic speedups becomes less meaningful. The key question is not about asymptotic behaviour, but rather about practical advantages for specific fixed-sized problems. 

Machine learning and optimization face similar considerations. While these fields have problems that are hard for classical computers in the worst case, practical instances often possess rich structures that classical heuristics can exploit. Analogously, many quantum heuristics have been proposed---some of which we expect to succeed, while others may fail. However, validating these approaches remains challenging due to the lack of appropriate quantum hardware. This makes it difficult to map out provable exponential speedups in commercially valuable applications, even in the fault-tolerant era.

The observation in \Cref{mythExp} about the current absence of proven exponential speedups with guaranteed commercial success is therefore correct. 
However, we anticipate that fault-tolerant quantum computers will enable empirically validated quantum heuristics---some of which may even lead to large provable (super)polynomial speed-ups for commercially relevant problems.
The critical question is whether quantum computers---near-term or fault-tolerant---can solve practically useful problems more effectively than classical approaches. Indeed, recent work \cite{childs2018toward, childs2021theory, rubin2024quantum} provides convincing arguments that quantum simulation of out-of-equilibrium dynamics could deliver substantial practical value for industrial applications. Success will ultimately be measured by our ability to address real-world challenges, regardless of whether the quantum advantage is polynomial or exponential.

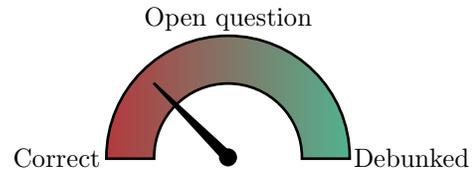
\begin{center}
    \gauge{-45}{Open question}
\end{center}

\begin{conclusion}
While proving exponential speedups for commercially valuable end-to-end applications remains mathematically challenging, it is reasonable to expect that quantum computers will eventually deliver meaningful speedups in practical problems.
\end{conclusion}

\section{Outlook}

Achieving practical quantum advantage is highly non-trivial, partly due to limitations posed by hardware noise and due to the fundamental hardness of circuit training. While these two are challenges that we focused on, at the same time, a true practical quantum advantage is further hindered by a fierce competition with advanced classical algorithms.

Still, we argue that there exist areas that are potentially fertile for demonstrating and practically exploring early quantum advantage. We expect the following three primary criteria to be crucial: First, we need applications where a modest circuit depth is sufficient. Second, in variational approaches to avoid barren plateaus and poor local minima, we need to employ problem-motivated Ans\"atze, classical precomputation, and appropriate cost functions and initial states. Finally, we need to focus on application areas with highest potential to overtake classical strategies, e.g., quantum simulation.

We believe that the late NISQ era with a few tens of thousands of gates could enable the earliest forms of practical quantum advantage. In this era, the main constraint is expected to be the limited circuit depth due to the exponential blow-up of the error mitigation overhead. One of the most tangible goals for near-term quantum advantage is simulating the dynamics of quantum many-body systems in regimes where the best known classical techniques fail. For this era the most promising direction could be discrete time evolutions, e.g., Floquet dynamics. Such time-evolutions have already been experimentally demonstrated in Refs.~\cite{Rosenberg_2024,Kim_2023,Fischer_2024}. For the case of continuous time evolutions a Trotterization overhead also has to be taken into account in the circuit depth. However, a range of recent developments alleviate the overhead associated with continuous time evolution, such as variational simulation of time  evolution~\cite{li2017VariaSim,kiumi2024te} or exploiting known differing energy scales in the Hamiltonian~\cite{Bosse24}, whereby the increase in circuit depth may be moderate.

As a next evolutionary stage, we anticipate that soon the first early fault tolerant machines will emerge which will enable gate errors of order of $10^{-6}-10^{-8}$ allowing relatively deeper circuits. Thus, these architectures will unlock more advanced algorithms such as phase estimation~\cite{kitaev1995QMeasAndAbelianStabilizer, lee2023EvidenceQuantAdvGroundState}, quantum signal processing~\cite{low2016HamSimQSignProc,gilyen2018QSingValTransf}, and Gibbs sampling~\cite{PRXQuantum.3.040305,chen2023QThermalStatePrep}. In the specific case of Gibbs samplers, one may speculate that some variants could enable efficient thermal state preparation at temperatures proportional to the hardware noise level, thus they might be applicable in the late NISQ and early fault tolerant eras. Although there are no known protocols with such guarantees yet, recently a similar noise resilience property was proven for a dissipative ground state preparation algorithm~\cite{cubitt2023dissipativegroundstatepreparation}, giving hope in designing novel quantum algorithms with natural error resilience.

In classical computing, Monte Carlo algorithms such as Gibbs sampling were also initially developed for physics simulation but are today used ubiquitously, often within a heuristic setting. Analogously, also various types of quantum heuristics have been proposed, however,  without any serious validation.
In the early fault-tolerant era, one will have the opportunity to test various quantum heuristics.
Through the selection and validation of good heuristic primitives, more complex solutions could be built; 
analogously to how in machine learning unexpected breakthroughs eventually materialized after a long journey of trial-and-error development, enabled by improved data sets and hardware --- despite discouraging early no-go results.

Thinking outside the box, we could also view quantum computers not only as tools for speeding up computations but also as devices for handling more efficiently quantum data. Here we would mention three potential use-cases.
In communication or key-distribution problems, quantum computers could be used to devise better quantum encodings that exploit non-local degrees of freedom to maximize the rate and to route information among different nodes in a quantum network.
Another approach could be driving coherently a quantum sensor with a quantum computer; this would in particular be useful for enhancing sensing of time-dependent fields \cite{allen2024sensing}. 
Finally, combining the above, we could foresee that distributed quantum computers could enhance sensing spatial- and time-varying fields.

Finally, we acknowledge that quantum computers need to be scalable and fully fault tolerant to entirely fulfil their societally transformative potential. Nevertheless, we emphasize that it may take a long evolution through the late NISQ and early fault-tolerant eras to achieve these long-term prospects, and, despite many technical challenges and fundamental limitations ahead, it is highly plausible that useful applications may be within reach before full fault tolerance is achieved.

\section*{Acknowledgments} 
\noindent This work was initiated by stimulating discussions at the Quantum Now Unconference, organised and sponsored by Algorithmiq, held from April 3-6, 2024. It was at this event that the concept for this perspective article was conceived and the initial draft of its content was created. The authors also thank the organisers of the Seeking Quantum Advantage (SEEQA)  conference that took place between September 2-5, 2024 in Oxford, where thought-provoking discussions led to further evolution of  article.

\bibliography{refs}

\begin{thebibliography}{97}%
\makeatletter
\providecommand \@ifxundefined [1]{%
 \@ifx{#1\undefined}
}%
\providecommand \@ifnum [1]{%
 \ifnum #1\expandafter \@firstoftwo
 \else \expandafter \@secondoftwo
 \fi
}%
\providecommand \@ifx [1]{%
 \ifx #1\expandafter \@firstoftwo
 \else \expandafter \@secondoftwo
 \fi
}%
\providecommand \natexlab [1]{#1}%
\providecommand \enquote  [1]{``#1''}%
\providecommand \bibnamefont  [1]{#1}%
\providecommand \bibfnamefont [1]{#1}%
\providecommand \citenamefont [1]{#1}%
\providecommand \href@noop [0]{\@secondoftwo}%
\providecommand \href [0]{\begingroup \@sanitize@url \@href}%
\providecommand \@href[1]{\@@startlink{#1}\@@href}%
\providecommand \@@href[1]{\endgroup#1\@@endlink}%
\providecommand \@sanitize@url [0]{\catcode `\\12\catcode `\$12\catcode `\&12\catcode `\#12\catcode `\^12\catcode `\_12\catcode `\%12\relax}%
\providecommand \@@startlink[1]{}%
\providecommand \@@endlink[0]{}%
\providecommand \url  [0]{\begingroup\@sanitize@url \@url }%
\providecommand \@url [1]{\endgroup\@href {#1}{\urlprefix }}%
\providecommand \urlprefix  [0]{URL }%
\providecommand \Eprint [0]{\href }%
\providecommand \doibase [0]{https://doi.org/}%
\providecommand \selectlanguage [0]{\@gobble}%
\providecommand \bibinfo  [0]{\@secondoftwo}%
\providecommand \bibfield  [0]{\@secondoftwo}%
\providecommand \translation [1]{[#1]}%
\providecommand \BibitemOpen [0]{}%
\providecommand \bibitemStop [0]{}%
\providecommand \bibitemNoStop [0]{.\EOS\space}%
\providecommand \EOS [0]{\spacefactor3000\relax}%
\providecommand \BibitemShut  [1]{\csname bibitem#1\endcsname}%
\let\auto@bib@innerbib\@empty
%</preamble>
\bibitem [{\citenamefont {Preskill}(2018)}]{preskill2018QuantCompNISQEra}%
  \BibitemOpen
  \bibfield  {author} {\bibinfo {author} {\bibfnamefont {J.}~\bibnamefont {Preskill}},\ }\bibfield  {title} {\bibinfo {title} {Quantum {C}omputing in the {NISQ} era and beyond},\ }\href {https://doi.org/10.22331/q-2018-08-06-79} {\bibfield  {journal} {\bibinfo  {journal} {{Quantum}}\ }\textbf {\bibinfo {volume} {2}},\ \bibinfo {pages} {79} (\bibinfo {year} {2018})}\BibitemShut {NoStop}%
\bibitem [{\citenamefont {Takagi}\ \emph {et~al.}(2022)\citenamefont {Takagi}, \citenamefont {Endo}, \citenamefont {Minagawa},\ and\ \citenamefont {Gu}}]{takagi2022fundamental}%
  \BibitemOpen
  \bibfield  {author} {\bibinfo {author} {\bibfnamefont {R.}~\bibnamefont {Takagi}}, \bibinfo {author} {\bibfnamefont {S.}~\bibnamefont {Endo}}, \bibinfo {author} {\bibfnamefont {S.}~\bibnamefont {Minagawa}},\ and\ \bibinfo {author} {\bibfnamefont {M.}~\bibnamefont {Gu}},\ }\bibfield  {title} {\bibinfo {title} {Fundamental limits of quantum error mitigation},\ }\href {https://doi.org/https://doi.org/10.1038/s41534-022-00618-z} {\bibfield  {journal} {\bibinfo  {journal} {npj Quantum Inf.}\ }\textbf {\bibinfo {volume} {8}},\ \bibinfo {pages} {114} (\bibinfo {year} {2022})}\BibitemShut {NoStop}%
\bibitem [{\citenamefont {Deshpande}\ \emph {et~al.}(2022)\citenamefont {Deshpande}, \citenamefont {Niroula}, \citenamefont {Shtanko}, \citenamefont {Gorshkov}, \citenamefont {Fefferman},\ and\ \citenamefont {Gullans}}]{deshpande2022tight}%
  \BibitemOpen
  \bibfield  {author} {\bibinfo {author} {\bibfnamefont {A.}~\bibnamefont {Deshpande}}, \bibinfo {author} {\bibfnamefont {P.}~\bibnamefont {Niroula}}, \bibinfo {author} {\bibfnamefont {O.}~\bibnamefont {Shtanko}}, \bibinfo {author} {\bibfnamefont {A.~V.}\ \bibnamefont {Gorshkov}}, \bibinfo {author} {\bibfnamefont {B.}~\bibnamefont {Fefferman}},\ and\ \bibinfo {author} {\bibfnamefont {M.~J.}\ \bibnamefont {Gullans}},\ }\bibfield  {title} {\bibinfo {title} {Tight bounds on the convergence of noisy random circuits to the uniform distribution},\ }\href {https://doi.org/https://doi.org/10.1103/PRXQuantum.3.040329} {\bibfield  {journal} {\bibinfo  {journal} {PRX Quantum}\ }\textbf {\bibinfo {volume} {3}},\ \bibinfo {pages} {040329} (\bibinfo {year} {2022})}\BibitemShut {NoStop}%
\bibitem [{\citenamefont {Takagi}\ \emph {et~al.}(2023)\citenamefont {Takagi}, \citenamefont {Tajima},\ and\ \citenamefont {Gu}}]{takagi2023universal}%
  \BibitemOpen
  \bibfield  {author} {\bibinfo {author} {\bibfnamefont {R.}~\bibnamefont {Takagi}}, \bibinfo {author} {\bibfnamefont {H.}~\bibnamefont {Tajima}},\ and\ \bibinfo {author} {\bibfnamefont {M.}~\bibnamefont {Gu}},\ }\bibfield  {title} {\bibinfo {title} {Universal sampling lower bounds for quantum error mitigation},\ }\href {https://doi.org/https://doi.org/10.1103/PhysRevLett.131.210602} {\bibfield  {journal} {\bibinfo  {journal} {Phys. Rev. Lett.}\ }\textbf {\bibinfo {volume} {131}},\ \bibinfo {pages} {210602} (\bibinfo {year} {2023})}\BibitemShut {NoStop}%
\bibitem [{\citenamefont {Tsubouchi}\ \emph {et~al.}(2023)\citenamefont {Tsubouchi}, \citenamefont {Sagawa},\ and\ \citenamefont {Yoshioka}}]{tsubouchi2023universal}%
  \BibitemOpen
  \bibfield  {author} {\bibinfo {author} {\bibfnamefont {K.}~\bibnamefont {Tsubouchi}}, \bibinfo {author} {\bibfnamefont {T.}~\bibnamefont {Sagawa}},\ and\ \bibinfo {author} {\bibfnamefont {N.}~\bibnamefont {Yoshioka}},\ }\bibfield  {title} {\bibinfo {title} {Universal cost bound of quantum error mitigation based on quantum estimation theory},\ }\href {https://doi.org/https://doi.org/10.1103/PhysRevLett.131.210601} {\bibfield  {journal} {\bibinfo  {journal} {Phys. Rev. Lett.}\ }\textbf {\bibinfo {volume} {131}},\ \bibinfo {pages} {210601} (\bibinfo {year} {2023})}\BibitemShut {NoStop}%
\bibitem [{\citenamefont {Quek}\ \emph {et~al.}(2024)\citenamefont {Quek}, \citenamefont {Stilck~França}, \citenamefont {Khatri}, \citenamefont {Meyer},\ and\ \citenamefont {Eisert}}]{quek2022exponentially}%
  \BibitemOpen
  \bibfield  {author} {\bibinfo {author} {\bibfnamefont {Y.}~\bibnamefont {Quek}}, \bibinfo {author} {\bibfnamefont {D.}~\bibnamefont {Stilck~França}}, \bibinfo {author} {\bibfnamefont {S.}~\bibnamefont {Khatri}}, \bibinfo {author} {\bibfnamefont {J.~J.}\ \bibnamefont {Meyer}},\ and\ \bibinfo {author} {\bibfnamefont {J.}~\bibnamefont {Eisert}},\ }\bibfield  {title} {\bibinfo {title} {Exponentially tighter bounds on limitations of quantum error mitigation},\ }\href {https://doi.org/10.1038/s41567-024-02536-7} {\bibfield  {journal} {\bibinfo  {journal} {Nat. Phys.}\ }\textbf {\bibinfo {volume} {20}},\ \bibinfo {pages} {1648–1658} (\bibinfo {year} {2024})}\BibitemShut {NoStop}%
\bibitem [{\citenamefont {Temme}\ \emph {et~al.}(2017)\citenamefont {Temme}, \citenamefont {Bravyi},\ and\ \citenamefont {Gambetta}}]{temme2017error}%
  \BibitemOpen
  \bibfield  {author} {\bibinfo {author} {\bibfnamefont {K.}~\bibnamefont {Temme}}, \bibinfo {author} {\bibfnamefont {S.}~\bibnamefont {Bravyi}},\ and\ \bibinfo {author} {\bibfnamefont {J.~M.}\ \bibnamefont {Gambetta}},\ }\bibfield  {title} {\bibinfo {title} {Error mitigation for short-depth quantum circuits},\ }\href {https://doi.org/10.1103/PhysRevLett.119.180509} {\bibfield  {journal} {\bibinfo  {journal} {Phys. Rev. Lett.}\ }\textbf {\bibinfo {volume} {119}},\ \bibinfo {pages} {180509} (\bibinfo {year} {2017})}\BibitemShut {NoStop}%
\bibitem [{\citenamefont {Endo}\ \emph {et~al.}(2018)\citenamefont {Endo}, \citenamefont {Benjamin},\ and\ \citenamefont {Li}}]{endo2018practical}%
  \BibitemOpen
  \bibfield  {author} {\bibinfo {author} {\bibfnamefont {S.}~\bibnamefont {Endo}}, \bibinfo {author} {\bibfnamefont {S.~C.}\ \bibnamefont {Benjamin}},\ and\ \bibinfo {author} {\bibfnamefont {Y.}~\bibnamefont {Li}},\ }\bibfield  {title} {\bibinfo {title} {Practical quantum error mitigation for near-future applications},\ }\href@noop {} {\bibfield  {journal} {\bibinfo  {journal} {Phys. Rev. X}\ }\textbf {\bibinfo {volume} {8}},\ \bibinfo {pages} {031027} (\bibinfo {year} {2018})}\BibitemShut {NoStop}%
\bibitem [{\citenamefont {Kim}\ \emph {et~al.}(2023{\natexlab{a}})\citenamefont {Kim}, \citenamefont {Wood}, \citenamefont {Yoder}, \citenamefont {Merkel}, \citenamefont {Gambetta}, \citenamefont {Temme},\ and\ \citenamefont {Kandala}}]{Kim_scalable_2023}%
  \BibitemOpen
  \bibfield  {author} {\bibinfo {author} {\bibfnamefont {Y.}~\bibnamefont {Kim}}, \bibinfo {author} {\bibfnamefont {C.~J.}\ \bibnamefont {Wood}}, \bibinfo {author} {\bibfnamefont {T.~J.}\ \bibnamefont {Yoder}}, \bibinfo {author} {\bibfnamefont {S.~T.}\ \bibnamefont {Merkel}}, \bibinfo {author} {\bibfnamefont {J.~M.}\ \bibnamefont {Gambetta}}, \bibinfo {author} {\bibfnamefont {K.}~\bibnamefont {Temme}},\ and\ \bibinfo {author} {\bibfnamefont {A.}~\bibnamefont {Kandala}},\ }\bibfield  {title} {\bibinfo {title} {Scalable error mitigation for noisy quantum circuits produces competitive expectation values},\ }\href {https://doi.org/10.1038/s41567-022-01914-3} {\bibfield  {journal} {\bibinfo  {journal} {Nat. Phys.}\ }\textbf {\bibinfo {volume} {19}},\ \bibinfo {pages} {752–759} (\bibinfo {year} {2023}{\natexlab{a}})}\BibitemShut {NoStop}%
\bibitem [{\citenamefont {Filippov}\ \emph {et~al.}(2023)\citenamefont {Filippov}, \citenamefont {Leahy}, \citenamefont {Rossi},\ and\ \citenamefont {García-Pérez}}]{filippov2023scalable}%
  \BibitemOpen
  \bibfield  {author} {\bibinfo {author} {\bibfnamefont {S.}~\bibnamefont {Filippov}}, \bibinfo {author} {\bibfnamefont {M.}~\bibnamefont {Leahy}}, \bibinfo {author} {\bibfnamefont {M.~A.~C.}\ \bibnamefont {Rossi}},\ and\ \bibinfo {author} {\bibfnamefont {G.}~\bibnamefont {García-Pérez}},\ }\href@noop {} {\bibinfo {title} {Scalable tensor-network error mitigation for near-term quantum computing}} (\bibinfo {year} {2023}),\ \Eprint {https://arxiv.org/abs/2307.11740} {arXiv:2307.11740 [quant-ph]} \BibitemShut {NoStop}%
\bibitem [{\citenamefont {Koczor}(2021)}]{PhysRevX.11.031057}%
  \BibitemOpen
  \bibfield  {author} {\bibinfo {author} {\bibfnamefont {B.}~\bibnamefont {Koczor}},\ }\bibfield  {title} {\bibinfo {title} {Exponential error suppression for near-term quantum devices},\ }\href {https://doi.org/10.1103/PhysRevX.11.031057} {\bibfield  {journal} {\bibinfo  {journal} {Phys. Rev. X}\ }\textbf {\bibinfo {volume} {11}},\ \bibinfo {pages} {031057} (\bibinfo {year} {2021})}\BibitemShut {NoStop}%
\bibitem [{\citenamefont {Huggins}\ \emph {et~al.}(2021)\citenamefont {Huggins}, \citenamefont {McArdle}, \citenamefont {O'Brien}, \citenamefont {Lee}, \citenamefont {Rubin}, \citenamefont {Boixo}, \citenamefont {Whaley}, \citenamefont {Babbush},\ and\ \citenamefont {McClean}}]{PhysRevX.11.041036}%
  \BibitemOpen
  \bibfield  {author} {\bibinfo {author} {\bibfnamefont {W.~J.}\ \bibnamefont {Huggins}}, \bibinfo {author} {\bibfnamefont {S.}~\bibnamefont {McArdle}}, \bibinfo {author} {\bibfnamefont {T.~E.}\ \bibnamefont {O'Brien}}, \bibinfo {author} {\bibfnamefont {J.}~\bibnamefont {Lee}}, \bibinfo {author} {\bibfnamefont {N.~C.}\ \bibnamefont {Rubin}}, \bibinfo {author} {\bibfnamefont {S.}~\bibnamefont {Boixo}}, \bibinfo {author} {\bibfnamefont {K.~B.}\ \bibnamefont {Whaley}}, \bibinfo {author} {\bibfnamefont {R.}~\bibnamefont {Babbush}},\ and\ \bibinfo {author} {\bibfnamefont {J.~R.}\ \bibnamefont {McClean}},\ }\bibfield  {title} {\bibinfo {title} {Virtual distillation for quantum error mitigation},\ }\href {https://doi.org/10.1103/PhysRevX.11.041036} {\bibfield  {journal} {\bibinfo  {journal} {Phys. Rev. X}\ }\textbf {\bibinfo {volume} {11}},\ \bibinfo {pages} {041036} (\bibinfo {year} {2021})}\BibitemShut {NoStop}%
\bibitem [{\citenamefont {Czarnik}\ \emph {et~al.}(2021)\citenamefont {Czarnik}, \citenamefont {Arrasmith}, \citenamefont {Coles},\ and\ \citenamefont {Cincio}}]{czarnik2021error}%
  \BibitemOpen
  \bibfield  {author} {\bibinfo {author} {\bibfnamefont {P.}~\bibnamefont {Czarnik}}, \bibinfo {author} {\bibfnamefont {A.}~\bibnamefont {Arrasmith}}, \bibinfo {author} {\bibfnamefont {P.~J.}\ \bibnamefont {Coles}},\ and\ \bibinfo {author} {\bibfnamefont {L.}~\bibnamefont {Cincio}},\ }\bibfield  {title} {\bibinfo {title} {Error mitigation with {C}lifford quantum-circuit data},\ }\href {https://doi.org/10.22331/q-2021-11-26-592} {\bibfield  {journal} {\bibinfo  {journal} {Quantum}\ }\textbf {\bibinfo {volume} {5}},\ \bibinfo {pages} {592} (\bibinfo {year} {2021})}\BibitemShut {NoStop}%
\bibitem [{\citenamefont {Maciejewski}\ \emph {et~al.}(2020)\citenamefont {Maciejewski}, \citenamefont {Zimbor{\'a}s},\ and\ \citenamefont {Oszmaniec}}]{maciejewski2020mitigation}%
  \BibitemOpen
  \bibfield  {author} {\bibinfo {author} {\bibfnamefont {F.~B.}\ \bibnamefont {Maciejewski}}, \bibinfo {author} {\bibfnamefont {Z.}~\bibnamefont {Zimbor{\'a}s}},\ and\ \bibinfo {author} {\bibfnamefont {M.}~\bibnamefont {Oszmaniec}},\ }\bibfield  {title} {\bibinfo {title} {Mitigation of readout noise in near-term quantum devices by classical post-processing based on detector tomography},\ }\href {https://doi.org/10.22331/q-2020-04-24-257} {\bibfield  {journal} {\bibinfo  {journal} {Quantum}\ }\textbf {\bibinfo {volume} {4}},\ \bibinfo {pages} {257} (\bibinfo {year} {2020})}\BibitemShut {NoStop}%
\bibitem [{\citenamefont {Van Den~Berg}\ \emph {et~al.}(2022)\citenamefont {Van Den~Berg}, \citenamefont {Minev},\ and\ \citenamefont {Temme}}]{van2022model}%
  \BibitemOpen
  \bibfield  {author} {\bibinfo {author} {\bibfnamefont {E.}~\bibnamefont {Van Den~Berg}}, \bibinfo {author} {\bibfnamefont {Z.~K.}\ \bibnamefont {Minev}},\ and\ \bibinfo {author} {\bibfnamefont {K.}~\bibnamefont {Temme}},\ }\bibfield  {title} {\bibinfo {title} {Model-free readout-error mitigation for quantum expectation values},\ }\href {https://link.aps.org/doi/10.1103/PhysRevA.105.032620} {\bibfield  {journal} {\bibinfo  {journal} {Phys. Rev. A}\ }\textbf {\bibinfo {volume} {105}},\ \bibinfo {pages} {032620} (\bibinfo {year} {2022})}\BibitemShut {NoStop}%
\bibitem [{\citenamefont {Cai}\ \emph {et~al.}(2023)\citenamefont {Cai}, \citenamefont {Babbush}, \citenamefont {Benjamin}, \citenamefont {Endo}, \citenamefont {Huggins}, \citenamefont {Li}, \citenamefont {McClean},\ and\ \citenamefont {O’Brien}}]{cai2023quantum}%
  \BibitemOpen
  \bibfield  {author} {\bibinfo {author} {\bibfnamefont {Z.}~\bibnamefont {Cai}}, \bibinfo {author} {\bibfnamefont {R.}~\bibnamefont {Babbush}}, \bibinfo {author} {\bibfnamefont {S.~C.}\ \bibnamefont {Benjamin}}, \bibinfo {author} {\bibfnamefont {S.}~\bibnamefont {Endo}}, \bibinfo {author} {\bibfnamefont {W.~J.}\ \bibnamefont {Huggins}}, \bibinfo {author} {\bibfnamefont {Y.}~\bibnamefont {Li}}, \bibinfo {author} {\bibfnamefont {J.~R.}\ \bibnamefont {McClean}},\ and\ \bibinfo {author} {\bibfnamefont {T.~E.}\ \bibnamefont {O’Brien}},\ }\bibfield  {title} {\bibinfo {title} {Quantum error mitigation},\ }\href {https://doi.org/10.1103/RevModPhys.95.045005} {\bibfield  {journal} {\bibinfo  {journal} {Rev. Mod. Phys.}\ }\textbf {\bibinfo {volume} {95}},\ \bibinfo {pages} {045005} (\bibinfo {year} {2023})}\BibitemShut {NoStop}%
\bibitem [{\citenamefont {Su}\ \emph {et~al.}(2021)\citenamefont {Su}, \citenamefont {Israel}, \citenamefont {Sharma}, \citenamefont {Qi}, \citenamefont {Dhand},\ and\ \citenamefont {Br{\'a}dler}}]{su2021error}%
  \BibitemOpen
  \bibfield  {author} {\bibinfo {author} {\bibfnamefont {D.}~\bibnamefont {Su}}, \bibinfo {author} {\bibfnamefont {R.}~\bibnamefont {Israel}}, \bibinfo {author} {\bibfnamefont {K.}~\bibnamefont {Sharma}}, \bibinfo {author} {\bibfnamefont {H.}~\bibnamefont {Qi}}, \bibinfo {author} {\bibfnamefont {I.}~\bibnamefont {Dhand}},\ and\ \bibinfo {author} {\bibfnamefont {K.}~\bibnamefont {Br{\'a}dler}},\ }\bibfield  {title} {\bibinfo {title} {Error mitigation on a near-term quantum photonic device},\ }\href {https://doi.org/10.22331/q-2021-05-04-452} {\bibfield  {journal} {\bibinfo  {journal} {Quantum}\ }\textbf {\bibinfo {volume} {5}},\ \bibinfo {pages} {452} (\bibinfo {year} {2021})}\BibitemShut {NoStop}%
\bibitem [{\citenamefont {Taylor}\ \emph {et~al.}(2024)\citenamefont {Taylor}, \citenamefont {Bressanini}, \citenamefont {Kwon},\ and\ \citenamefont {Kim}}]{taylor2024quantum}%
  \BibitemOpen
  \bibfield  {author} {\bibinfo {author} {\bibfnamefont {A.}~\bibnamefont {Taylor}}, \bibinfo {author} {\bibfnamefont {G.}~\bibnamefont {Bressanini}}, \bibinfo {author} {\bibfnamefont {H.}~\bibnamefont {Kwon}},\ and\ \bibinfo {author} {\bibfnamefont {M.}~\bibnamefont {Kim}},\ }\bibfield  {title} {\bibinfo {title} {Quantum error cancellation in photonic systems: Undoing photon losses},\ }\href {https://doi.org/10.1103/PhysRevA.110.022622} {\bibfield  {journal} {\bibinfo  {journal} {Phys. Rev. A}\ }\textbf {\bibinfo {volume} {110}},\ \bibinfo {pages} {022622} (\bibinfo {year} {2024})}\BibitemShut {NoStop}%
\bibitem [{\citenamefont {Faurby}\ \emph {et~al.}(2024)\citenamefont {Faurby}, \citenamefont {Carosini}, \citenamefont {Cao}, \citenamefont {Sund}, \citenamefont {Hansen}, \citenamefont {Giorgino}, \citenamefont {Villadsen}, \citenamefont {van~den Hoven}, \citenamefont {Lodahl}, \citenamefont {Paesani}, \citenamefont {Loredo},\ and\ \citenamefont {Walther}}]{faurby2024purifying}%
  \BibitemOpen
  \bibfield  {author} {\bibinfo {author} {\bibfnamefont {C.~F.~D.}\ \bibnamefont {Faurby}}, \bibinfo {author} {\bibfnamefont {L.}~\bibnamefont {Carosini}}, \bibinfo {author} {\bibfnamefont {H.}~\bibnamefont {Cao}}, \bibinfo {author} {\bibfnamefont {P.~I.}\ \bibnamefont {Sund}}, \bibinfo {author} {\bibfnamefont {L.~M.}\ \bibnamefont {Hansen}}, \bibinfo {author} {\bibfnamefont {F.}~\bibnamefont {Giorgino}}, \bibinfo {author} {\bibfnamefont {A.~B.}\ \bibnamefont {Villadsen}}, \bibinfo {author} {\bibfnamefont {S.~N.}\ \bibnamefont {van~den Hoven}}, \bibinfo {author} {\bibfnamefont {P.}~\bibnamefont {Lodahl}}, \bibinfo {author} {\bibfnamefont {S.}~\bibnamefont {Paesani}}, \bibinfo {author} {\bibfnamefont {J.~C.}\ \bibnamefont {Loredo}},\ and\ \bibinfo {author} {\bibfnamefont {P.}~\bibnamefont {Walther}},\ }\bibfield  {title} {\bibinfo {title} {Purifying photon indistinguishability through quantum interference},\ }\href {https://doi.org/10.1103/PhysRevLett.133.033604} {\bibfield  {journal} {\bibinfo  {journal}
  {Phys. Rev. Lett.}\ }\textbf {\bibinfo {volume} {133}},\ \bibinfo {pages} {033604} (\bibinfo {year} {2024})}\BibitemShut {NoStop}%
\bibitem [{\citenamefont {Acharya}\ \emph {et~al.}(2024)\citenamefont {Acharya} \emph {et~al.}}]{Acharya_2024_N}%
  \BibitemOpen
  \bibfield  {author} {\bibinfo {author} {\bibfnamefont {R.}~\bibnamefont {Acharya}} \emph {et~al.},\ }\bibfield  {title} {\bibinfo {title} {Quantum error correction below the surface code threshold},\ }\href {https://doi.org/10.1038/s41586-024-08449-y} {\bibfield  {journal} {\bibinfo  {journal} {Nature}\ } (\bibinfo {year} {2024})}\BibitemShut {NoStop}%
\bibitem [{IBM(2024)}]{IBM_QPUs}%
  \BibitemOpen
  \href {https://quantum.ibm.com/services/resources} {\bibinfo {title} {Quantum processing units}} (\bibinfo {year} {2024}),\ \bibinfo {note} {accessed on December 9, 2024}\BibitemShut {NoStop}%
\bibitem [{\citenamefont {Evered}\ \emph {et~al.}(2023)\citenamefont {Evered}, \citenamefont {Bluvstein}, \citenamefont {Kalinowski}, \citenamefont {Ebadi}, \citenamefont {Manovitz}, \citenamefont {Zhou}, \citenamefont {Li}, \citenamefont {Geim}, \citenamefont {Wang}, \citenamefont {Maskara}, \citenamefont {Levine}, \citenamefont {Semeghini}, \citenamefont {Greiner}, \citenamefont {Vuletić},\ and\ \citenamefont {Lukin}}]{evered2023high}%
  \BibitemOpen
  \bibfield  {author} {\bibinfo {author} {\bibfnamefont {S.~J.}\ \bibnamefont {Evered}}, \bibinfo {author} {\bibfnamefont {D.}~\bibnamefont {Bluvstein}}, \bibinfo {author} {\bibfnamefont {M.}~\bibnamefont {Kalinowski}}, \bibinfo {author} {\bibfnamefont {S.}~\bibnamefont {Ebadi}}, \bibinfo {author} {\bibfnamefont {T.}~\bibnamefont {Manovitz}}, \bibinfo {author} {\bibfnamefont {H.}~\bibnamefont {Zhou}}, \bibinfo {author} {\bibfnamefont {S.~H.}\ \bibnamefont {Li}}, \bibinfo {author} {\bibfnamefont {A.~A.}\ \bibnamefont {Geim}}, \bibinfo {author} {\bibfnamefont {T.~T.}\ \bibnamefont {Wang}}, \bibinfo {author} {\bibfnamefont {N.}~\bibnamefont {Maskara}}, \bibinfo {author} {\bibfnamefont {H.}~\bibnamefont {Levine}}, \bibinfo {author} {\bibfnamefont {G.}~\bibnamefont {Semeghini}}, \bibinfo {author} {\bibfnamefont {M.}~\bibnamefont {Greiner}}, \bibinfo {author} {\bibfnamefont {V.}~\bibnamefont {Vuletić}},\ and\ \bibinfo {author} {\bibfnamefont {M.~D.}\ \bibnamefont {Lukin}},\ }\bibfield  {title} {\bibinfo {title}
  {High-fidelity parallel entangling gates on a neutral-atom quantum computer},\ }\href {https://doi.org/10.1038/s41586-023-06481-y} {\bibfield  {journal} {\bibinfo  {journal} {Nature}\ }\textbf {\bibinfo {volume} {622}},\ \bibinfo {pages} {268} (\bibinfo {year} {2023})}\BibitemShut {NoStop}%
\bibitem [{\citenamefont {Chen}\ \emph {et~al.}(2024)\citenamefont {Chen}, \citenamefont {Nielsen}, \citenamefont {Ebert}, \citenamefont {Inlek}, \citenamefont {Wright}, \citenamefont {Chaplin}, \citenamefont {Maksymov}, \citenamefont {P{\'{a}}ez}, \citenamefont {Poudel}, \citenamefont {Maunz},\ and\ \citenamefont {Gamble}}]{chen2023benchmarking}%
  \BibitemOpen
  \bibfield  {author} {\bibinfo {author} {\bibfnamefont {J.-S.}\ \bibnamefont {Chen}}, \bibinfo {author} {\bibfnamefont {E.}~\bibnamefont {Nielsen}}, \bibinfo {author} {\bibfnamefont {M.}~\bibnamefont {Ebert}}, \bibinfo {author} {\bibfnamefont {V.}~\bibnamefont {Inlek}}, \bibinfo {author} {\bibfnamefont {K.}~\bibnamefont {Wright}}, \bibinfo {author} {\bibfnamefont {V.}~\bibnamefont {Chaplin}}, \bibinfo {author} {\bibfnamefont {A.}~\bibnamefont {Maksymov}}, \bibinfo {author} {\bibfnamefont {E.}~\bibnamefont {P{\'{a}}ez}}, \bibinfo {author} {\bibfnamefont {A.}~\bibnamefont {Poudel}}, \bibinfo {author} {\bibfnamefont {P.}~\bibnamefont {Maunz}},\ and\ \bibinfo {author} {\bibfnamefont {J.}~\bibnamefont {Gamble}},\ }\bibfield  {title} {\bibinfo {title} {Benchmarking a trapped-ion quantum computer with 30 qubits},\ }\href {https://doi.org/10.22331/q-2024-11-07-1516} {\bibfield  {journal} {\bibinfo  {journal} {{Quantum}}\ }\textbf {\bibinfo {volume} {8}},\ \bibinfo {pages} {1516} (\bibinfo {year} {2024})}\BibitemShut
  {NoStop}%
\bibitem [{\citenamefont {Khan}\ and\ \citenamefont {Strabley}(2024)}]{H2}%
  \BibitemOpen
  \bibfield  {author} {\bibinfo {author} {\bibfnamefont {I.}~\bibnamefont {Khan}}\ and\ \bibinfo {author} {\bibfnamefont {J.}~\bibnamefont {Strabley}},\ }\href {https://www.quantinuum.com/blog/quantinuum-extends-its-significant-lead-in-quantum-computing-achieving-historic-milestones-for-hardware-fidelity-and-quantum-volume} {\bibinfo {title} {{Q}uantinuum {B}log}} (\bibinfo {year} {2024})\BibitemShut {NoStop}%
\bibitem [{\citenamefont {Löschnauer}\ \emph {et~al.}(2024)\citenamefont {Löschnauer}, \citenamefont {Toba}, \citenamefont {Hughes}, \citenamefont {King}, \citenamefont {Weber}, \citenamefont {Srinivas}, \citenamefont {Matt}, \citenamefont {Nourshargh}, \citenamefont {Allcock}, \citenamefont {Ballance}, \citenamefont {Matthiesen}, \citenamefont {Malinowski},\ and\ \citenamefont {Harty}}]{loschnauer2024scalable}%
  \BibitemOpen
  \bibfield  {author} {\bibinfo {author} {\bibfnamefont {C.~M.}\ \bibnamefont {Löschnauer}}, \bibinfo {author} {\bibfnamefont {J.~M.}\ \bibnamefont {Toba}}, \bibinfo {author} {\bibfnamefont {A.~C.}\ \bibnamefont {Hughes}}, \bibinfo {author} {\bibfnamefont {S.~A.}\ \bibnamefont {King}}, \bibinfo {author} {\bibfnamefont {M.~A.}\ \bibnamefont {Weber}}, \bibinfo {author} {\bibfnamefont {R.}~\bibnamefont {Srinivas}}, \bibinfo {author} {\bibfnamefont {R.}~\bibnamefont {Matt}}, \bibinfo {author} {\bibfnamefont {R.}~\bibnamefont {Nourshargh}}, \bibinfo {author} {\bibfnamefont {D.~T.~C.}\ \bibnamefont {Allcock}}, \bibinfo {author} {\bibfnamefont {C.~J.}\ \bibnamefont {Ballance}}, \bibinfo {author} {\bibfnamefont {C.}~\bibnamefont {Matthiesen}}, \bibinfo {author} {\bibfnamefont {M.}~\bibnamefont {Malinowski}},\ and\ \bibinfo {author} {\bibfnamefont {T.~P.}\ \bibnamefont {Harty}},\ }\href@noop {} {\bibinfo {title} {Scalable, high-fidelity all-electronic control of trapped-ion qubits}} (\bibinfo {year} {2024}),\ \Eprint
  {https://arxiv.org/abs/2407.07694} {arXiv:2407.07694 [quant-ph]} \BibitemShut {NoStop}%
\bibitem [{\citenamefont {M{\"u}ller-Hermes}\ \emph {et~al.}(2016)\citenamefont {M{\"u}ller-Hermes}, \citenamefont {Stilck~Fran{\c{c}}a},\ and\ \citenamefont {Wolf}}]{muller2016relative}%
  \BibitemOpen
  \bibfield  {author} {\bibinfo {author} {\bibfnamefont {A.}~\bibnamefont {M{\"u}ller-Hermes}}, \bibinfo {author} {\bibfnamefont {D.}~\bibnamefont {Stilck~Fran{\c{c}}a}},\ and\ \bibinfo {author} {\bibfnamefont {M.~M.}\ \bibnamefont {Wolf}},\ }\bibfield  {title} {\bibinfo {title} {Relative entropy convergence for depolarizing channels},\ }\href {https://doi.org/10.1063/1.4939560} {\bibfield  {journal} {\bibinfo  {journal} {J. Math. Phys.}\ }\textbf {\bibinfo {volume} {57}},\ \bibinfo {pages} {022202} (\bibinfo {year} {2016})}\BibitemShut {NoStop}%
\bibitem [{\citenamefont {Stilck~Fran{\c{c}}a}\ and\ \citenamefont {Garcia-Patron}(2021)}]{stilck2021limitations}%
  \BibitemOpen
  \bibfield  {author} {\bibinfo {author} {\bibfnamefont {D.}~\bibnamefont {Stilck~Fran{\c{c}}a}}\ and\ \bibinfo {author} {\bibfnamefont {R.}~\bibnamefont {Garcia-Patron}},\ }\bibfield  {title} {\bibinfo {title} {Limitations of optimization algorithms on noisy quantum devices},\ }\href {https://doi.org/10.1038/s41567-021-01356-3} {\bibfield  {journal} {\bibinfo  {journal} {Nat. Phys.}\ }\textbf {\bibinfo {volume} {17}},\ \bibinfo {pages} {1221} (\bibinfo {year} {2021})}\BibitemShut {NoStop}%
\bibitem [{\citenamefont {Filippov}\ \emph {et~al.}(2024)\citenamefont {Filippov}, \citenamefont {Maniscalco},\ and\ \citenamefont {García-Pérez}}]{filippov2024scalability}%
  \BibitemOpen
  \bibfield  {author} {\bibinfo {author} {\bibfnamefont {S.~N.}\ \bibnamefont {Filippov}}, \bibinfo {author} {\bibfnamefont {S.}~\bibnamefont {Maniscalco}},\ and\ \bibinfo {author} {\bibfnamefont {G.}~\bibnamefont {García-Pérez}},\ }\href@noop {} {\bibinfo {title} {Scalability of quantum error mitigation techniques: from utility to advantage}} (\bibinfo {year} {2024}),\ \Eprint {https://arxiv.org/abs/2403.13542} {arXiv:2403.13542 [quant-ph]} \BibitemShut {NoStop}%
\bibitem [{\citenamefont {Proctor}\ \emph {et~al.}(2024)\citenamefont {Proctor}, \citenamefont {Young}, \citenamefont {Baczewski},\ and\ \citenamefont {Blume-Kohout}}]{Proctor_2024}%
  \BibitemOpen
  \bibfield  {author} {\bibinfo {author} {\bibfnamefont {T.}~\bibnamefont {Proctor}}, \bibinfo {author} {\bibfnamefont {K.}~\bibnamefont {Young}}, \bibinfo {author} {\bibfnamefont {A.~D.}\ \bibnamefont {Baczewski}},\ and\ \bibinfo {author} {\bibfnamefont {R.}~\bibnamefont {Blume-Kohout}},\ }\href@noop {} {\bibinfo {title} {Benchmarking quantum computers}} (\bibinfo {year} {2024}),\ \Eprint {https://arxiv.org/abs/2407.08828} {arXiv:2407.08828 [quant-ph]} \BibitemShut {NoStop}%
\bibitem [{\citenamefont {Arute}\ \emph {et~al.}(2019)\citenamefont {Arute} \emph {et~al.}}]{arute2019quantum}%
  \BibitemOpen
  \bibfield  {author} {\bibinfo {author} {\bibfnamefont {F.}~\bibnamefont {Arute}} \emph {et~al.},\ }\bibfield  {title} {\bibinfo {title} {Quantum supremacy using a programmable superconducting processor},\ }\href {https://doi.org/10.1038/s41586-019-1666-5} {\bibfield  {journal} {\bibinfo  {journal} {Nature}\ }\textbf {\bibinfo {volume} {574}},\ \bibinfo {pages} {505} (\bibinfo {year} {2019})}\BibitemShut {NoStop}%
\bibitem [{\citenamefont {Zhong}\ \emph {et~al.}(2020)\citenamefont {Zhong} \emph {et~al.}}]{Zhong_2020}%
  \BibitemOpen
  \bibfield  {author} {\bibinfo {author} {\bibfnamefont {H.-S.}\ \bibnamefont {Zhong}} \emph {et~al.},\ }\bibfield  {title} {\bibinfo {title} {Quantum computational advantage using photons},\ }\href {https://doi.org/10.1126/science.abe8770} {\bibfield  {journal} {\bibinfo  {journal} {Science}\ }\textbf {\bibinfo {volume} {370}},\ \bibinfo {pages} {1460–1463} (\bibinfo {year} {2020})}\BibitemShut {NoStop}%
\bibitem [{\citenamefont {Morvan}\ \emph {et~al.}(2024)\citenamefont {Morvan} \emph {et~al.}}]{Morvan_2024}%
  \BibitemOpen
  \bibfield  {author} {\bibinfo {author} {\bibfnamefont {A.}~\bibnamefont {Morvan}} \emph {et~al.},\ }\bibfield  {title} {\bibinfo {title} {Phase transitions in random circuit sampling},\ }\href {https://doi.org/10.1038/s41586-024-07998-6} {\bibfield  {journal} {\bibinfo  {journal} {Nature}\ }\textbf {\bibinfo {volume} {634}},\ \bibinfo {pages} {328–333} (\bibinfo {year} {2024})}\BibitemShut {NoStop}%
\bibitem [{\citenamefont {Gambetta}\ \emph {et~al.}(2021)\citenamefont {Gambetta}, \citenamefont {Javadi-Abhari}, \citenamefont {Johnson}, \citenamefont {Jurcevic}, \citenamefont {Paik},\ and\ \citenamefont {Wack}}]{gambetta2021driving}%
  \BibitemOpen
  \bibfield  {author} {\bibinfo {author} {\bibfnamefont {J.}~\bibnamefont {Gambetta}}, \bibinfo {author} {\bibfnamefont {A.}~\bibnamefont {Javadi-Abhari}}, \bibinfo {author} {\bibfnamefont {B.}~\bibnamefont {Johnson}}, \bibinfo {author} {\bibfnamefont {P.}~\bibnamefont {Jurcevic}}, \bibinfo {author} {\bibfnamefont {H.}~\bibnamefont {Paik}},\ and\ \bibinfo {author} {\bibfnamefont {A.}~\bibnamefont {Wack}},\ }\bibfield  {title} {\bibinfo {title} {Driving quantum performance: \-more qubits, higher quantum volume, and now a proper\- measure of speed},\ }\href {https://www.ibm.com/quantum/blog/circuit-layer-operations-per-second} {\bibfield  {journal} {\bibinfo  {journal} {IBM Blog}\ } (\bibinfo {year} {2021})}\BibitemShut {NoStop}%
\bibitem [{\citenamefont {Smith}\ \emph {et~al.}(2019)\citenamefont {Smith}, \citenamefont {Kim}, \citenamefont {Pollmann},\ and\ \citenamefont {Knolle}}]{Smith2019}%
  \BibitemOpen
  \bibfield  {author} {\bibinfo {author} {\bibfnamefont {A.}~\bibnamefont {Smith}}, \bibinfo {author} {\bibfnamefont {M.~S.}\ \bibnamefont {Kim}}, \bibinfo {author} {\bibfnamefont {F.}~\bibnamefont {Pollmann}},\ and\ \bibinfo {author} {\bibfnamefont {J.}~\bibnamefont {Knolle}},\ }\bibfield  {title} {\bibinfo {title} {Simulating quantum many-body dynamics on a current digital quantum computer},\ }\href {https://doi.org/10.1038/s41534-019-0217-0} {\bibfield  {journal} {\bibinfo  {journal} {npj Quantum Inf.}\ }\textbf {\bibinfo {volume} {5}},\ \bibinfo {pages} {106} (\bibinfo {year} {2019})}\BibitemShut {NoStop}%
\bibitem [{\citenamefont {Clinton}\ \emph {et~al.}(2021)\citenamefont {Clinton}, \citenamefont {Bausch},\ and\ \citenamefont {Cubitt}}]{Clinton2021}%
  \BibitemOpen
  \bibfield  {author} {\bibinfo {author} {\bibfnamefont {L.}~\bibnamefont {Clinton}}, \bibinfo {author} {\bibfnamefont {J.}~\bibnamefont {Bausch}},\ and\ \bibinfo {author} {\bibfnamefont {T.}~\bibnamefont {Cubitt}},\ }\bibfield  {title} {\bibinfo {title} {Hamiltonian simulation algorithms for near-term quantum hardware},\ }\href {https://doi.org/10.1038/s41467-021-25196-0} {\bibfield  {journal} {\bibinfo  {journal} {Nat. Commun.}\ }\textbf {\bibinfo {volume} {12}},\ \bibinfo {pages} {4989} (\bibinfo {year} {2021})}\BibitemShut {NoStop}%
\bibitem [{\citenamefont {Clinton}\ \emph {et~al.}(2024)\citenamefont {Clinton}, \citenamefont {Cubitt}, \citenamefont {Flynn}, \citenamefont {Gambetta}, \citenamefont {Klassen}, \citenamefont {Montanaro}, \citenamefont {Piddock}, \citenamefont {Santos},\ and\ \citenamefont {Sheridan}}]{Clinton2024}%
  \BibitemOpen
  \bibfield  {author} {\bibinfo {author} {\bibfnamefont {L.}~\bibnamefont {Clinton}}, \bibinfo {author} {\bibfnamefont {T.}~\bibnamefont {Cubitt}}, \bibinfo {author} {\bibfnamefont {B.}~\bibnamefont {Flynn}}, \bibinfo {author} {\bibfnamefont {F.~M.}\ \bibnamefont {Gambetta}}, \bibinfo {author} {\bibfnamefont {J.}~\bibnamefont {Klassen}}, \bibinfo {author} {\bibfnamefont {A.}~\bibnamefont {Montanaro}}, \bibinfo {author} {\bibfnamefont {S.}~\bibnamefont {Piddock}}, \bibinfo {author} {\bibfnamefont {R.~A.}\ \bibnamefont {Santos}},\ and\ \bibinfo {author} {\bibfnamefont {E.}~\bibnamefont {Sheridan}},\ }\bibfield  {title} {\bibinfo {title} {Towards near-term quantum simulation of materials},\ }\href {https://doi.org/10.1038/s41467-023-43479-6} {\bibfield  {journal} {\bibinfo  {journal} {Nat. Commun.}\ }\textbf {\bibinfo {volume} {15}},\ \bibinfo {pages} {211} (\bibinfo {year} {2024})}\BibitemShut {NoStop}%
\bibitem [{\citenamefont {Piveteau}\ \emph {et~al.}(2021)\citenamefont {Piveteau}, \citenamefont {Sutter}, \citenamefont {Bravyi}, \citenamefont {Gambetta},\ and\ \citenamefont {Temme}}]{piveteauErrorMitigationUniversal2021}%
  \BibitemOpen
  \bibfield  {author} {\bibinfo {author} {\bibfnamefont {C.}~\bibnamefont {Piveteau}}, \bibinfo {author} {\bibfnamefont {D.}~\bibnamefont {Sutter}}, \bibinfo {author} {\bibfnamefont {S.}~\bibnamefont {Bravyi}}, \bibinfo {author} {\bibfnamefont {J.~M.}\ \bibnamefont {Gambetta}},\ and\ \bibinfo {author} {\bibfnamefont {K.}~\bibnamefont {Temme}},\ }\bibfield  {title} {\bibinfo {title} {Error {{Mitigation}} for {{Universal Gates}} on {{Encoded Qubits}}},\ }\href {https://doi.org/10.1103/PhysRevLett.127.200505} {\bibfield  {journal} {\bibinfo  {journal} {Phys. Rev. Lett.}\ }\textbf {\bibinfo {volume} {127}},\ \bibinfo {pages} {200505} (\bibinfo {year} {2021})}\BibitemShut {NoStop}%
\bibitem [{\citenamefont {Suzuki}\ \emph {et~al.}(2022)\citenamefont {Suzuki}, \citenamefont {Endo}, \citenamefont {Fujii},\ and\ \citenamefont {Tokunaga}}]{suzuki2022quantum}%
  \BibitemOpen
  \bibfield  {author} {\bibinfo {author} {\bibfnamefont {Y.}~\bibnamefont {Suzuki}}, \bibinfo {author} {\bibfnamefont {S.}~\bibnamefont {Endo}}, \bibinfo {author} {\bibfnamefont {K.}~\bibnamefont {Fujii}},\ and\ \bibinfo {author} {\bibfnamefont {Y.}~\bibnamefont {Tokunaga}},\ }\bibfield  {title} {\bibinfo {title} {Quantum error mitigation as a universal error reduction technique: Applications from the {NISQ} to the fault-tolerant quantum computing eras},\ }\href {https://doi.org/10.1103/PRXQuantum.3.010345} {\bibfield  {journal} {\bibinfo  {journal} {PRX Quantum}\ }\textbf {\bibinfo {volume} {3}},\ \bibinfo {pages} {010345} (\bibinfo {year} {2022})}\BibitemShut {NoStop}%
\bibitem [{\citenamefont {Wahl}\ \emph {et~al.}(2023)\citenamefont {Wahl}, \citenamefont {Mari}, \citenamefont {Shammah}, \citenamefont {Zeng},\ and\ \citenamefont {Ravi}}]{wahl2023zero}%
  \BibitemOpen
  \bibfield  {author} {\bibinfo {author} {\bibfnamefont {M.~A.}\ \bibnamefont {Wahl}}, \bibinfo {author} {\bibfnamefont {A.}~\bibnamefont {Mari}}, \bibinfo {author} {\bibfnamefont {N.}~\bibnamefont {Shammah}}, \bibinfo {author} {\bibfnamefont {W.~J.}\ \bibnamefont {Zeng}},\ and\ \bibinfo {author} {\bibfnamefont {G.~S.}\ \bibnamefont {Ravi}},\ }\bibfield  {title} {\bibinfo {title} {Zero noise extrapolation on logical qubits by scaling the error correction code distance},\ }in\ \href@noop {} {\emph {\bibinfo {booktitle} {2023 IEEE International Conference on Quantum Computing and Engineering (QCE)}}}\ (\bibinfo  {publisher} {IEEE},\ \bibinfo {year} {2023})\ p.\ \bibinfo {pages} {888–897}\BibitemShut {NoStop}%
\bibitem [{\citenamefont {Koczor}\ \emph {et~al.}(2024)\citenamefont {Koczor}, \citenamefont {Morton},\ and\ \citenamefont {Benjamin}}]{PhysRevLett.132.130602}%
  \BibitemOpen
  \bibfield  {author} {\bibinfo {author} {\bibfnamefont {B.}~\bibnamefont {Koczor}}, \bibinfo {author} {\bibfnamefont {J.~J.~L.}\ \bibnamefont {Morton}},\ and\ \bibinfo {author} {\bibfnamefont {S.~C.}\ \bibnamefont {Benjamin}},\ }\bibfield  {title} {\bibinfo {title} {Probabilistic interpolation of quantum rotation angles},\ }\href {https://doi.org/10.1103/PhysRevLett.132.130602} {\bibfield  {journal} {\bibinfo  {journal} {Phys. Rev. Lett.}\ }\textbf {\bibinfo {volume} {132}},\ \bibinfo {pages} {130602} (\bibinfo {year} {2024})}\BibitemShut {NoStop}%
\bibitem [{\citenamefont {Koczor}(2024)}]{koczor2024sparse}%
  \BibitemOpen
  \bibfield  {author} {\bibinfo {author} {\bibfnamefont {B.}~\bibnamefont {Koczor}},\ }\bibfield  {title} {\bibinfo {title} {Sparse probabilistic synthesis of quantum operations},\ }\href {https://doi.org/10.1103/PRXQuantum.5.040352} {\bibfield  {journal} {\bibinfo  {journal} {PRX Quantum}\ }\textbf {\bibinfo {volume} {5}},\ \bibinfo {pages} {040352} (\bibinfo {year} {2024})}\BibitemShut {NoStop}%
\bibitem [{\citenamefont {Kliuchnikov}\ \emph {et~al.}(2023)\citenamefont {Kliuchnikov}, \citenamefont {Lauter}, \citenamefont {Minko}, \citenamefont {Paetznick},\ and\ \citenamefont {Petit}}]{kliuchnikov2023shorter}%
  \BibitemOpen
  \bibfield  {author} {\bibinfo {author} {\bibfnamefont {V.}~\bibnamefont {Kliuchnikov}}, \bibinfo {author} {\bibfnamefont {K.}~\bibnamefont {Lauter}}, \bibinfo {author} {\bibfnamefont {R.}~\bibnamefont {Minko}}, \bibinfo {author} {\bibfnamefont {A.}~\bibnamefont {Paetznick}},\ and\ \bibinfo {author} {\bibfnamefont {C.}~\bibnamefont {Petit}},\ }\bibfield  {title} {\bibinfo {title} {Shorter quantum circuits via single-qubit gate approximation},\ }\href {https://doi.org/10.22331/q-2023-12-18-1208} {\bibfield  {journal} {\bibinfo  {journal} {Quantum}\ }\textbf {\bibinfo {volume} {7}},\ \bibinfo {pages} {1208} (\bibinfo {year} {2023})}\BibitemShut {NoStop}%
\bibitem [{\citenamefont {Endo}\ \emph {et~al.}(2019)\citenamefont {Endo}, \citenamefont {Zhao}, \citenamefont {Li}, \citenamefont {Benjamin},\ and\ \citenamefont {Yuan}}]{PhysRevA.99.012334}%
  \BibitemOpen
  \bibfield  {author} {\bibinfo {author} {\bibfnamefont {S.}~\bibnamefont {Endo}}, \bibinfo {author} {\bibfnamefont {Q.}~\bibnamefont {Zhao}}, \bibinfo {author} {\bibfnamefont {Y.}~\bibnamefont {Li}}, \bibinfo {author} {\bibfnamefont {S.}~\bibnamefont {Benjamin}},\ and\ \bibinfo {author} {\bibfnamefont {X.}~\bibnamefont {Yuan}},\ }\bibfield  {title} {\bibinfo {title} {Mitigating algorithmic errors in a {H}amiltonian simulation},\ }\href {https://doi.org/10.1103/PhysRevA.99.012334} {\bibfield  {journal} {\bibinfo  {journal} {Phys. Rev. A}\ }\textbf {\bibinfo {volume} {99}},\ \bibinfo {pages} {012334} (\bibinfo {year} {2019})}\BibitemShut {NoStop}%
\bibitem [{\citenamefont {Rendon}\ \emph {et~al.}(2024)\citenamefont {Rendon}, \citenamefont {Watkins},\ and\ \citenamefont {Wiebe}}]{rendon2024improved}%
  \BibitemOpen
  \bibfield  {author} {\bibinfo {author} {\bibfnamefont {G.}~\bibnamefont {Rendon}}, \bibinfo {author} {\bibfnamefont {J.}~\bibnamefont {Watkins}},\ and\ \bibinfo {author} {\bibfnamefont {N.}~\bibnamefont {Wiebe}},\ }\bibfield  {title} {\bibinfo {title} {Improved accuracy for {T}rotter simulations using {C}hebyshev interpolation},\ }\href {https://doi.org/10.22331/q-2024-02-26-1266} {\bibfield  {journal} {\bibinfo  {journal} {Quantum}\ }\textbf {\bibinfo {volume} {8}},\ \bibinfo {pages} {1266} (\bibinfo {year} {2024})}\BibitemShut {NoStop}%
\bibitem [{\citenamefont {Chan}\ \emph {et~al.}(2024)\citenamefont {Chan}, \citenamefont {Meister}, \citenamefont {Goh},\ and\ \citenamefont {Koczor}}]{chan2022algorithmic}%
  \BibitemOpen
  \bibfield  {author} {\bibinfo {author} {\bibfnamefont {H.~H.~S.}\ \bibnamefont {Chan}}, \bibinfo {author} {\bibfnamefont {R.}~\bibnamefont {Meister}}, \bibinfo {author} {\bibfnamefont {M.~L.}\ \bibnamefont {Goh}},\ and\ \bibinfo {author} {\bibfnamefont {B.}~\bibnamefont {Koczor}},\ }\href@noop {} {\bibinfo {title} {Algorithmic shadow spectroscopy}} (\bibinfo {year} {2024}),\ \Eprint {https://arxiv.org/abs/2212.11036} {arXiv:2212.11036 [quant-ph]} \BibitemShut {NoStop}%
\bibitem [{\citenamefont {Kiumi}\ and\ \citenamefont {Koczor}(2024)}]{kiumi2024te}%
  \BibitemOpen
  \bibfield  {author} {\bibinfo {author} {\bibfnamefont {C.}~\bibnamefont {Kiumi}}\ and\ \bibinfo {author} {\bibfnamefont {B.}~\bibnamefont {Koczor}},\ }\href@noop {} {\bibinfo {title} {{TE-PAI}: Exact time evolution by sampling random circuits}} (\bibinfo {year} {2024}),\ \Eprint {https://arxiv.org/abs/2410.16850} {arXiv:2410.16850 [quant-ph]} \BibitemShut {NoStop}%
\bibitem [{\citenamefont {Liu}\ \emph {et~al.}(2024)\citenamefont {Liu}, \citenamefont {Zhang}, \citenamefont {Fei},\ and\ \citenamefont {Cai}}]{liu2024virtualchannelpurification}%
  \BibitemOpen
  \bibfield  {author} {\bibinfo {author} {\bibfnamefont {Z.}~\bibnamefont {Liu}}, \bibinfo {author} {\bibfnamefont {X.}~\bibnamefont {Zhang}}, \bibinfo {author} {\bibfnamefont {Y.-Y.}\ \bibnamefont {Fei}},\ and\ \bibinfo {author} {\bibfnamefont {Z.}~\bibnamefont {Cai}},\ }\href@noop {} {\bibinfo {title} {Virtual channel purification}} (\bibinfo {year} {2024}),\ \Eprint {https://arxiv.org/abs/2402.07866} {arXiv:2402.07866 [quant-ph]} \BibitemShut {NoStop}%
\bibitem [{\citenamefont {Carleo}\ \emph {et~al.}(2019)\citenamefont {Carleo}, \citenamefont {Cirac}, \citenamefont {Cranmer}, \citenamefont {Daudet}, \citenamefont {Schuld}, \citenamefont {Tishby}, \citenamefont {Vogt-Maranto},\ and\ \citenamefont {Zdeborov\'a}}]{carleo2019machine}%
  \BibitemOpen
  \bibfield  {author} {\bibinfo {author} {\bibfnamefont {G.}~\bibnamefont {Carleo}}, \bibinfo {author} {\bibfnamefont {I.}~\bibnamefont {Cirac}}, \bibinfo {author} {\bibfnamefont {K.}~\bibnamefont {Cranmer}}, \bibinfo {author} {\bibfnamefont {L.}~\bibnamefont {Daudet}}, \bibinfo {author} {\bibfnamefont {M.}~\bibnamefont {Schuld}}, \bibinfo {author} {\bibfnamefont {N.}~\bibnamefont {Tishby}}, \bibinfo {author} {\bibfnamefont {L.}~\bibnamefont {Vogt-Maranto}},\ and\ \bibinfo {author} {\bibfnamefont {L.}~\bibnamefont {Zdeborov\'a}},\ }\bibfield  {title} {\bibinfo {title} {Machine learning and the physical sciences},\ }\href {https://doi.org/10.1103/RevModPhys.91.045002} {\bibfield  {journal} {\bibinfo  {journal} {Rev. Mod. Phys.}\ }\textbf {\bibinfo {volume} {91}},\ \bibinfo {pages} {045002} (\bibinfo {year} {2019})}\BibitemShut {NoStop}%
\bibitem [{\citenamefont {Cerezo}\ \emph {et~al.}(2021)\citenamefont {Cerezo}, \citenamefont {Arrasmith}, \citenamefont {Babbush}, \citenamefont {Benjamin}, \citenamefont {Endo}, \citenamefont {Fujii}, \citenamefont {McClean}, \citenamefont {Mitarai}, \citenamefont {Yuan}, \citenamefont {Cincio},\ and\ \citenamefont {Coles}}]{cerezo2020variationalreview}%
  \BibitemOpen
  \bibfield  {author} {\bibinfo {author} {\bibfnamefont {M.}~\bibnamefont {Cerezo}}, \bibinfo {author} {\bibfnamefont {A.}~\bibnamefont {Arrasmith}}, \bibinfo {author} {\bibfnamefont {R.}~\bibnamefont {Babbush}}, \bibinfo {author} {\bibfnamefont {S.~C.}\ \bibnamefont {Benjamin}}, \bibinfo {author} {\bibfnamefont {S.}~\bibnamefont {Endo}}, \bibinfo {author} {\bibfnamefont {K.}~\bibnamefont {Fujii}}, \bibinfo {author} {\bibfnamefont {J.~R.}\ \bibnamefont {McClean}}, \bibinfo {author} {\bibfnamefont {K.}~\bibnamefont {Mitarai}}, \bibinfo {author} {\bibfnamefont {X.}~\bibnamefont {Yuan}}, \bibinfo {author} {\bibfnamefont {L.}~\bibnamefont {Cincio}},\ and\ \bibinfo {author} {\bibfnamefont {P.~J.}\ \bibnamefont {Coles}},\ }\bibfield  {title} {\bibinfo {title} {Variational quantum algorithms},\ }\href {https://doi.org/10.1038/s42254-021-00348-9} {\bibfield  {journal} {\bibinfo  {journal} {Nat. Rev. Phys.}\ }\textbf {\bibinfo {volume} {3}},\ \bibinfo {pages} {625–644} (\bibinfo {year} {2021})}\BibitemShut {NoStop}%
\bibitem [{\citenamefont {Endo}\ \emph {et~al.}(2021)\citenamefont {Endo}, \citenamefont {Cai}, \citenamefont {Benjamin},\ and\ \citenamefont {Yuan}}]{endo2021hybrid}%
  \BibitemOpen
  \bibfield  {author} {\bibinfo {author} {\bibfnamefont {S.}~\bibnamefont {Endo}}, \bibinfo {author} {\bibfnamefont {Z.}~\bibnamefont {Cai}}, \bibinfo {author} {\bibfnamefont {S.~C.}\ \bibnamefont {Benjamin}},\ and\ \bibinfo {author} {\bibfnamefont {X.}~\bibnamefont {Yuan}},\ }\bibfield  {title} {\bibinfo {title} {Hybrid quantum-classical algorithms and quantum error mitigation},\ }\href {https://doi.org/10.7566/JPSJ.90.032001} {\bibfield  {journal} {\bibinfo  {journal} {J. Phys. Soc. Jpn.}\ }\textbf {\bibinfo {volume} {90}},\ \bibinfo {pages} {032001} (\bibinfo {year} {2021})}\BibitemShut {NoStop}%
\bibitem [{\citenamefont {Biamonte}\ \emph {et~al.}(2017)\citenamefont {Biamonte}, \citenamefont {Wittek}, \citenamefont {Pancotti}, \citenamefont {Rebentrost}, \citenamefont {Wiebe},\ and\ \citenamefont {Lloyd}}]{biamonte2017quantum}%
  \BibitemOpen
  \bibfield  {author} {\bibinfo {author} {\bibfnamefont {J.}~\bibnamefont {Biamonte}}, \bibinfo {author} {\bibfnamefont {P.}~\bibnamefont {Wittek}}, \bibinfo {author} {\bibfnamefont {N.}~\bibnamefont {Pancotti}}, \bibinfo {author} {\bibfnamefont {P.}~\bibnamefont {Rebentrost}}, \bibinfo {author} {\bibfnamefont {N.}~\bibnamefont {Wiebe}},\ and\ \bibinfo {author} {\bibfnamefont {S.}~\bibnamefont {Lloyd}},\ }\bibfield  {title} {\bibinfo {title} {Quantum machine learning},\ }\href {https://doi.org/10.1038/nature23474} {\bibfield  {journal} {\bibinfo  {journal} {Nature}\ }\textbf {\bibinfo {volume} {549}},\ \bibinfo {pages} {195} (\bibinfo {year} {2017})}\BibitemShut {NoStop}%
\bibitem [{\citenamefont {Schuld}\ and\ \citenamefont {Petruccione}(2018)}]{schuld2018supervised}%
  \BibitemOpen
  \bibfield  {author} {\bibinfo {author} {\bibfnamefont {M.}~\bibnamefont {Schuld}}\ and\ \bibinfo {author} {\bibfnamefont {F.}~\bibnamefont {Petruccione}},\ }\href@noop {} {\emph {\bibinfo {title} {Supervised learning with quantum computers}}}\ (\bibinfo  {publisher} {Springer},\ \bibinfo {year} {2018})\BibitemShut {NoStop}%
\bibitem [{\citenamefont {Cerezo}\ \emph {et~al.}(2022)\citenamefont {Cerezo}, \citenamefont {Verdon}, \citenamefont {Huang}, \citenamefont {Cincio},\ and\ \citenamefont {Coles}}]{cerezo2022challenges}%
  \BibitemOpen
  \bibfield  {author} {\bibinfo {author} {\bibfnamefont {M.}~\bibnamefont {Cerezo}}, \bibinfo {author} {\bibfnamefont {G.}~\bibnamefont {Verdon}}, \bibinfo {author} {\bibfnamefont {H.-Y.}\ \bibnamefont {Huang}}, \bibinfo {author} {\bibfnamefont {L.}~\bibnamefont {Cincio}},\ and\ \bibinfo {author} {\bibfnamefont {P.~J.}\ \bibnamefont {Coles}},\ }\bibfield  {title} {\bibinfo {title} {Challenges and opportunities in quantum machine learning},\ }\href {https://doi.org/10.1038/s43588-022-00311-3} {\bibfield  {journal} {\bibinfo  {journal} {Nat. Comput. Sci.}\ }\textbf {\bibinfo {volume} {2}},\ \bibinfo {pages} {567–576} (\bibinfo {year} {2022})}\BibitemShut {NoStop}%
\bibitem [{\citenamefont {Bittel}\ and\ \citenamefont {Kliesch}(2021)}]{bittel2021training}%
  \BibitemOpen
  \bibfield  {author} {\bibinfo {author} {\bibfnamefont {L.}~\bibnamefont {Bittel}}\ and\ \bibinfo {author} {\bibfnamefont {M.}~\bibnamefont {Kliesch}},\ }\bibfield  {title} {\bibinfo {title} {Training variational quantum algorithms is {NP}-hard},\ }\href {https://doi.org/10.1103/PhysRevLett.127.120502} {\bibfield  {journal} {\bibinfo  {journal} {Phys. Rev. Lett.}\ }\textbf {\bibinfo {volume} {127}},\ \bibinfo {pages} {120502} (\bibinfo {year} {2021})}\BibitemShut {NoStop}%
\bibitem [{\citenamefont {Fontana}\ \emph {et~al.}(2022)\citenamefont {Fontana}, \citenamefont {Cerezo}, \citenamefont {Arrasmith}, \citenamefont {Rungger},\ and\ \citenamefont {Coles}}]{fontana2022nontrivial}%
  \BibitemOpen
  \bibfield  {author} {\bibinfo {author} {\bibfnamefont {E.}~\bibnamefont {Fontana}}, \bibinfo {author} {\bibfnamefont {M.}~\bibnamefont {Cerezo}}, \bibinfo {author} {\bibfnamefont {A.}~\bibnamefont {Arrasmith}}, \bibinfo {author} {\bibfnamefont {I.}~\bibnamefont {Rungger}},\ and\ \bibinfo {author} {\bibfnamefont {P.~J.}\ \bibnamefont {Coles}},\ }\bibfield  {title} {\bibinfo {title} {Non-trivial symmetries in quantum landscapes and their resilience to quantum noise},\ }\href {https://doi.org/10.22331/q-2022-09-15-804} {\bibfield  {journal} {\bibinfo  {journal} {Quantum}\ }\textbf {\bibinfo {volume} {6}},\ \bibinfo {pages} {804} (\bibinfo {year} {2022})}\BibitemShut {NoStop}%
\bibitem [{\citenamefont {Anschuetz}\ and\ \citenamefont {Kiani}(2022)}]{anschuetz2022beyond}%
  \BibitemOpen
  \bibfield  {author} {\bibinfo {author} {\bibfnamefont {E.~R.}\ \bibnamefont {Anschuetz}}\ and\ \bibinfo {author} {\bibfnamefont {B.~T.}\ \bibnamefont {Kiani}},\ }\bibfield  {title} {\bibinfo {title} {Quantum variational algorithms are swamped with traps},\ }\href {https://doi.org/10.1038/s41467-022-35364-5} {\bibfield  {journal} {\bibinfo  {journal} {Nat. Commun.}\ }\textbf {\bibinfo {volume} {13}},\ \bibinfo {pages} {7760} (\bibinfo {year} {2022})}\BibitemShut {NoStop}%
\bibitem [{\citenamefont {Anschuetz}(2022)}]{anschuetz2021critical}%
  \BibitemOpen
  \bibfield  {author} {\bibinfo {author} {\bibfnamefont {E.~R.}\ \bibnamefont {Anschuetz}},\ }\bibfield  {title} {\bibinfo {title} {Critical points in quantum generative models},\ }in\ \href {https://openreview.net/forum?id=2f1z55GVQN} {\emph {\bibinfo {booktitle} {International Conference on Learning Representations}}}\ (\bibinfo {year} {2022})\BibitemShut {NoStop}%
\bibitem [{\citenamefont {Larocca}\ \emph {et~al.}(2022)\citenamefont {Larocca}, \citenamefont {Czarnik}, \citenamefont {Sharma}, \citenamefont {Muraleedharan}, \citenamefont {Coles},\ and\ \citenamefont {Cerezo}}]{larocca2021diagnosing}%
  \BibitemOpen
  \bibfield  {author} {\bibinfo {author} {\bibfnamefont {M.}~\bibnamefont {Larocca}}, \bibinfo {author} {\bibfnamefont {P.}~\bibnamefont {Czarnik}}, \bibinfo {author} {\bibfnamefont {K.}~\bibnamefont {Sharma}}, \bibinfo {author} {\bibfnamefont {G.}~\bibnamefont {Muraleedharan}}, \bibinfo {author} {\bibfnamefont {P.~J.}\ \bibnamefont {Coles}},\ and\ \bibinfo {author} {\bibfnamefont {M.}~\bibnamefont {Cerezo}},\ }\bibfield  {title} {\bibinfo {title} {Diagnosing {B}arren {P}lateaus with {T}ools from {Q}uantum {O}ptimal {C}ontrol},\ }\href {https://doi.org/10.22331/q-2022-09-29-824} {\bibfield  {journal} {\bibinfo  {journal} {{Quantum}}\ }\textbf {\bibinfo {volume} {6}},\ \bibinfo {pages} {824} (\bibinfo {year} {2022})}\BibitemShut {NoStop}%
\bibitem [{\citenamefont {Tikku}\ and\ \citenamefont {Kim}(2022)}]{tikku2022circuit}%
  \BibitemOpen
  \bibfield  {author} {\bibinfo {author} {\bibfnamefont {A.}~\bibnamefont {Tikku}}\ and\ \bibinfo {author} {\bibfnamefont {I.~H.}\ \bibnamefont {Kim}},\ }\href@noop {} {\bibinfo {title} {Circuit depth versus energy in topologically ordered systems}} (\bibinfo {year} {2022}),\ \Eprint {https://arxiv.org/abs/2210.06796} {arXiv:2210.06796 [quant-ph]} \BibitemShut {NoStop}%
\bibitem [{\citenamefont {McClean}\ \emph {et~al.}(2018)\citenamefont {McClean}, \citenamefont {Boixo}, \citenamefont {Smelyanskiy}, \citenamefont {Babbush},\ and\ \citenamefont {Neven}}]{mcclean2018barren}%
  \BibitemOpen
  \bibfield  {author} {\bibinfo {author} {\bibfnamefont {J.~R.}\ \bibnamefont {McClean}}, \bibinfo {author} {\bibfnamefont {S.}~\bibnamefont {Boixo}}, \bibinfo {author} {\bibfnamefont {V.~N.}\ \bibnamefont {Smelyanskiy}}, \bibinfo {author} {\bibfnamefont {R.}~\bibnamefont {Babbush}},\ and\ \bibinfo {author} {\bibfnamefont {H.}~\bibnamefont {Neven}},\ }\bibfield  {title} {\bibinfo {title} {Barren plateaus in quantum neural network training landscapes},\ }\href {https://doi.org/10.1038/s41467-018-07090-4} {\bibfield  {journal} {\bibinfo  {journal} {Nat. Commun.}\ }\textbf {\bibinfo {volume} {9}},\ \bibinfo {pages} {4812} (\bibinfo {year} {2018})}\BibitemShut {NoStop}%
\bibitem [{\citenamefont {Larocca}\ \emph {et~al.}(2024)\citenamefont {Larocca}, \citenamefont {Thanasilp}, \citenamefont {Wang}, \citenamefont {Sharma}, \citenamefont {Biamonte}, \citenamefont {Coles}, \citenamefont {Cincio}, \citenamefont {McClean}, \citenamefont {Holmes},\ and\ \citenamefont {Cerezo}}]{larocca2024review}%
  \BibitemOpen
  \bibfield  {author} {\bibinfo {author} {\bibfnamefont {M.}~\bibnamefont {Larocca}}, \bibinfo {author} {\bibfnamefont {S.}~\bibnamefont {Thanasilp}}, \bibinfo {author} {\bibfnamefont {S.}~\bibnamefont {Wang}}, \bibinfo {author} {\bibfnamefont {K.}~\bibnamefont {Sharma}}, \bibinfo {author} {\bibfnamefont {J.}~\bibnamefont {Biamonte}}, \bibinfo {author} {\bibfnamefont {P.~J.}\ \bibnamefont {Coles}}, \bibinfo {author} {\bibfnamefont {L.}~\bibnamefont {Cincio}}, \bibinfo {author} {\bibfnamefont {J.~R.}\ \bibnamefont {McClean}}, \bibinfo {author} {\bibfnamefont {Z.}~\bibnamefont {Holmes}},\ and\ \bibinfo {author} {\bibfnamefont {M.}~\bibnamefont {Cerezo}},\ }\href@noop {} {\bibinfo {title} {A review of barren plateaus in variational quantum computing}} (\bibinfo {year} {2024}),\ \Eprint {https://arxiv.org/abs/2405.00781} {arXiv:2405.00781 [quant-ph]} \BibitemShut {NoStop}%
\bibitem [{\citenamefont {Letcher}\ \emph {et~al.}(2024)\citenamefont {Letcher}, \citenamefont {Woerner},\ and\ \citenamefont {Zoufal}}]{letcher2023tight}%
  \BibitemOpen
  \bibfield  {author} {\bibinfo {author} {\bibfnamefont {A.}~\bibnamefont {Letcher}}, \bibinfo {author} {\bibfnamefont {S.}~\bibnamefont {Woerner}},\ and\ \bibinfo {author} {\bibfnamefont {C.}~\bibnamefont {Zoufal}},\ }\bibfield  {title} {\bibinfo {title} {Tight and efficient gradient bounds for parameterized quantum circuits},\ }\href {https://doi.org/10.22331/q-2024-09-25-1484} {\bibfield  {journal} {\bibinfo  {journal} {Quantum}\ }\textbf {\bibinfo {volume} {8}},\ \bibinfo {pages} {1484} (\bibinfo {year} {2024})}\BibitemShut {NoStop}%
\bibitem [{\citenamefont {Cerezo}\ \emph {et~al.}(2024)\citenamefont {Cerezo}, \citenamefont {Larocca}, \citenamefont {García-Martín}, \citenamefont {Diaz}, \citenamefont {Braccia}, \citenamefont {Fontana}, \citenamefont {Rudolph}, \citenamefont {Bermejo}, \citenamefont {Ijaz}, \citenamefont {Thanasilp}, \citenamefont {Anschuetz},\ and\ \citenamefont {Holmes}}]{cerezo2023does}%
  \BibitemOpen
  \bibfield  {author} {\bibinfo {author} {\bibfnamefont {M.}~\bibnamefont {Cerezo}}, \bibinfo {author} {\bibfnamefont {M.}~\bibnamefont {Larocca}}, \bibinfo {author} {\bibfnamefont {D.}~\bibnamefont {García-Martín}}, \bibinfo {author} {\bibfnamefont {N.~L.}\ \bibnamefont {Diaz}}, \bibinfo {author} {\bibfnamefont {P.}~\bibnamefont {Braccia}}, \bibinfo {author} {\bibfnamefont {E.}~\bibnamefont {Fontana}}, \bibinfo {author} {\bibfnamefont {M.~S.}\ \bibnamefont {Rudolph}}, \bibinfo {author} {\bibfnamefont {P.}~\bibnamefont {Bermejo}}, \bibinfo {author} {\bibfnamefont {A.}~\bibnamefont {Ijaz}}, \bibinfo {author} {\bibfnamefont {S.}~\bibnamefont {Thanasilp}}, \bibinfo {author} {\bibfnamefont {E.~R.}\ \bibnamefont {Anschuetz}},\ and\ \bibinfo {author} {\bibfnamefont {Z.}~\bibnamefont {Holmes}},\ }\href@noop {} {\bibinfo {title} {Does provable absence of barren plateaus imply classical simulability? {O}r, why we need to rethink variational quantum computing}} (\bibinfo {year} {2024}),\ \Eprint
  {https://arxiv.org/abs/2312.09121} {arXiv:2312.09121 [quant-ph]} \BibitemShut {NoStop}%
\bibitem [{\citenamefont {Bermejo}\ \emph {et~al.}(2024)\citenamefont {Bermejo}, \citenamefont {Braccia}, \citenamefont {Rudolph}, \citenamefont {Holmes}, \citenamefont {Cincio},\ and\ \citenamefont {Cerezo}}]{bermejo2024quantum}%
  \BibitemOpen
  \bibfield  {author} {\bibinfo {author} {\bibfnamefont {P.}~\bibnamefont {Bermejo}}, \bibinfo {author} {\bibfnamefont {P.}~\bibnamefont {Braccia}}, \bibinfo {author} {\bibfnamefont {M.~S.}\ \bibnamefont {Rudolph}}, \bibinfo {author} {\bibfnamefont {Z.}~\bibnamefont {Holmes}}, \bibinfo {author} {\bibfnamefont {L.}~\bibnamefont {Cincio}},\ and\ \bibinfo {author} {\bibfnamefont {M.}~\bibnamefont {Cerezo}},\ }\href@noop {} {\bibinfo {title} {Quantum convolutional neural networks are (effectively) classically simulable}} (\bibinfo {year} {2024}),\ \Eprint {https://arxiv.org/abs/2408.12739} {arXiv:2408.12739 [quant-ph]} \BibitemShut {NoStop}%
\bibitem [{\citenamefont {Khosrojerdi}\ \emph {et~al.}(2024)\citenamefont {Khosrojerdi}, \citenamefont {Pereira}, \citenamefont {Cuccoli},\ and\ \citenamefont {Banchi}}]{khosrojerdi2024learning}%
  \BibitemOpen
  \bibfield  {author} {\bibinfo {author} {\bibfnamefont {M.}~\bibnamefont {Khosrojerdi}}, \bibinfo {author} {\bibfnamefont {J.~L.}\ \bibnamefont {Pereira}}, \bibinfo {author} {\bibfnamefont {A.}~\bibnamefont {Cuccoli}},\ and\ \bibinfo {author} {\bibfnamefont {L.}~\bibnamefont {Banchi}},\ }\href@noop {} {\bibinfo {title} {Learning to classify quantum phases of matter with a few measurements}} (\bibinfo {year} {2024}),\ \Eprint {https://arxiv.org/abs/2409.05188} {arXiv:2409.05188 [quant-ph]} \BibitemShut {NoStop}%
\bibitem [{\citenamefont {Jerbi}\ \emph {et~al.}(2024)\citenamefont {Jerbi}, \citenamefont {Gyurik}, \citenamefont {Marshall}, \citenamefont {Molteni},\ and\ \citenamefont {Dunjko}}]{jerbi2024shadows}%
  \BibitemOpen
  \bibfield  {author} {\bibinfo {author} {\bibfnamefont {S.}~\bibnamefont {Jerbi}}, \bibinfo {author} {\bibfnamefont {C.}~\bibnamefont {Gyurik}}, \bibinfo {author} {\bibfnamefont {S.~C.}\ \bibnamefont {Marshall}}, \bibinfo {author} {\bibfnamefont {R.}~\bibnamefont {Molteni}},\ and\ \bibinfo {author} {\bibfnamefont {V.}~\bibnamefont {Dunjko}},\ }\bibfield  {title} {\bibinfo {title} {Shadows of quantum machine learning},\ }\href {https://doi.org/10.1038/s41467-024-49877-8} {\bibfield  {journal} {\bibinfo  {journal} {Nat. Commun.}\ }\textbf {\bibinfo {volume} {15}},\ \bibinfo {pages} {5676} (\bibinfo {year} {2024})}\BibitemShut {NoStop}%
\bibitem [{\citenamefont {Peruzzo}\ \emph {et~al.}(2014)\citenamefont {Peruzzo}, \citenamefont {McClean}, \citenamefont {Shadbolt}, \citenamefont {Yung}, \citenamefont {Zhou}, \citenamefont {Love}, \citenamefont {Aspuru-Guzik},\ and\ \citenamefont {O’Brien}}]{peruzzo2014variational}%
  \BibitemOpen
  \bibfield  {author} {\bibinfo {author} {\bibfnamefont {A.}~\bibnamefont {Peruzzo}}, \bibinfo {author} {\bibfnamefont {J.}~\bibnamefont {McClean}}, \bibinfo {author} {\bibfnamefont {P.}~\bibnamefont {Shadbolt}}, \bibinfo {author} {\bibfnamefont {M.-H.}\ \bibnamefont {Yung}}, \bibinfo {author} {\bibfnamefont {X.-Q.}\ \bibnamefont {Zhou}}, \bibinfo {author} {\bibfnamefont {P.~J.}\ \bibnamefont {Love}}, \bibinfo {author} {\bibfnamefont {A.}~\bibnamefont {Aspuru-Guzik}},\ and\ \bibinfo {author} {\bibfnamefont {J.~L.}\ \bibnamefont {O’Brien}},\ }\bibfield  {title} {\bibinfo {title} {A variational eigenvalue solver on a photonic quantum processor},\ }\href {https://doi.org/10.1038/ncomms5213} {\bibfield  {journal} {\bibinfo  {journal} {Nat. Commun.}\ }\textbf {\bibinfo {volume} {5}},\ \bibinfo {pages} {4213} (\bibinfo {year} {2014})}\BibitemShut {NoStop}%
\bibitem [{\citenamefont {Wecker}\ \emph {et~al.}(2015)\citenamefont {Wecker}, \citenamefont {Hastings},\ and\ \citenamefont {Troyer}}]{wecker2015progress}%
  \BibitemOpen
  \bibfield  {author} {\bibinfo {author} {\bibfnamefont {D.}~\bibnamefont {Wecker}}, \bibinfo {author} {\bibfnamefont {M.~B.}\ \bibnamefont {Hastings}},\ and\ \bibinfo {author} {\bibfnamefont {M.}~\bibnamefont {Troyer}},\ }\bibfield  {title} {\bibinfo {title} {Progress towards practical quantum variational algorithms},\ }\href {https://doi.org/10.1103/PhysRevA.92.042303} {\bibfield  {journal} {\bibinfo  {journal} {Phys. Rev. A}\ }\textbf {\bibinfo {volume} {92}},\ \bibinfo {pages} {042303} (\bibinfo {year} {2015})}\BibitemShut {NoStop}%
\bibitem [{\citenamefont {Hadfield}\ \emph {et~al.}(2019)\citenamefont {Hadfield}, \citenamefont {Wang}, \citenamefont {O'Gorman}, \citenamefont {Rieffel}, \citenamefont {Venturelli},\ and\ \citenamefont {Biswas}}]{hadfield2019quantum}%
  \BibitemOpen
  \bibfield  {author} {\bibinfo {author} {\bibfnamefont {S.}~\bibnamefont {Hadfield}}, \bibinfo {author} {\bibfnamefont {Z.}~\bibnamefont {Wang}}, \bibinfo {author} {\bibfnamefont {B.}~\bibnamefont {O'Gorman}}, \bibinfo {author} {\bibfnamefont {E.~G.}\ \bibnamefont {Rieffel}}, \bibinfo {author} {\bibfnamefont {D.}~\bibnamefont {Venturelli}},\ and\ \bibinfo {author} {\bibfnamefont {R.}~\bibnamefont {Biswas}},\ }\bibfield  {title} {\bibinfo {title} {From the quantum approximate optimization algorithm to a quantum alternating operator ansatz},\ }\href {https://doi.org/10.3390/a12020034} {\bibfield  {journal} {\bibinfo  {journal} {Algorithms}\ }\textbf {\bibinfo {volume} {12}},\ \bibinfo {pages} {34} (\bibinfo {year} {2019})}\BibitemShut {NoStop}%
\bibitem [{\citenamefont {Farhi}\ \emph {et~al.}(2014)\citenamefont {Farhi}, \citenamefont {Goldstone},\ and\ \citenamefont {Gutmann}}]{farhi2014quantum}%
  \BibitemOpen
  \bibfield  {author} {\bibinfo {author} {\bibfnamefont {E.}~\bibnamefont {Farhi}}, \bibinfo {author} {\bibfnamefont {J.}~\bibnamefont {Goldstone}},\ and\ \bibinfo {author} {\bibfnamefont {S.}~\bibnamefont {Gutmann}},\ }\href@noop {} {\bibinfo {title} {A quantum approximate optimization algorithm}} (\bibinfo {year} {2014}),\ \Eprint {https://arxiv.org/abs/1411.4028} {arXiv:1411.4028 [quant-ph]} \BibitemShut {NoStop}%
\bibitem [{\citenamefont {Schuld}\ \emph {et~al.}(2021)\citenamefont {Schuld}, \citenamefont {Sweke},\ and\ \citenamefont {Meyer}}]{schuld2021effect}%
  \BibitemOpen
  \bibfield  {author} {\bibinfo {author} {\bibfnamefont {M.}~\bibnamefont {Schuld}}, \bibinfo {author} {\bibfnamefont {R.}~\bibnamefont {Sweke}},\ and\ \bibinfo {author} {\bibfnamefont {J.~J.}\ \bibnamefont {Meyer}},\ }\bibfield  {title} {\bibinfo {title} {Effect of data encoding on the expressive power of variational quantum-machine-learning models},\ }\href {https://doi.org/10.1103/PhysRevA.103.032430} {\bibfield  {journal} {\bibinfo  {journal} {Phys. Rev. A}\ }\textbf {\bibinfo {volume} {103}},\ \bibinfo {pages} {032430} (\bibinfo {year} {2021})}\BibitemShut {NoStop}%
\bibitem [{\citenamefont {Larocca}\ \emph {et~al.}(2023)\citenamefont {Larocca}, \citenamefont {Ju}, \citenamefont {Garc{\'\i}a-Mart{\'\i}n}, \citenamefont {Coles},\ and\ \citenamefont {Cerezo}}]{larocca2023theory}%
  \BibitemOpen
  \bibfield  {author} {\bibinfo {author} {\bibfnamefont {M.}~\bibnamefont {Larocca}}, \bibinfo {author} {\bibfnamefont {N.}~\bibnamefont {Ju}}, \bibinfo {author} {\bibfnamefont {D.}~\bibnamefont {Garc{\'\i}a-Mart{\'\i}n}}, \bibinfo {author} {\bibfnamefont {P.~J.}\ \bibnamefont {Coles}},\ and\ \bibinfo {author} {\bibfnamefont {M.}~\bibnamefont {Cerezo}},\ }\bibfield  {title} {\bibinfo {title} {Theory of overparametrization in quantum neural networks},\ }\href {https://doi.org/10.1038/s43588-023-00467-6} {\bibfield  {journal} {\bibinfo  {journal} {Nat. Comput. Sci.}\ }\textbf {\bibinfo {volume} {3}},\ \bibinfo {pages} {542} (\bibinfo {year} {2023})}\BibitemShut {NoStop}%
\bibitem [{\citenamefont {Abbas}\ \emph {et~al.}(2023)\citenamefont {Abbas}, \citenamefont {King}, \citenamefont {Huang}, \citenamefont {Huggins}, \citenamefont {Movassagh}, \citenamefont {Gilboa},\ and\ \citenamefont {McClean}}]{abbas2024quantum}%
  \BibitemOpen
  \bibfield  {author} {\bibinfo {author} {\bibfnamefont {A.}~\bibnamefont {Abbas}}, \bibinfo {author} {\bibfnamefont {R.}~\bibnamefont {King}}, \bibinfo {author} {\bibfnamefont {H.-Y.}\ \bibnamefont {Huang}}, \bibinfo {author} {\bibfnamefont {W.~J.}\ \bibnamefont {Huggins}}, \bibinfo {author} {\bibfnamefont {R.}~\bibnamefont {Movassagh}}, \bibinfo {author} {\bibfnamefont {D.}~\bibnamefont {Gilboa}},\ and\ \bibinfo {author} {\bibfnamefont {J.~R.}\ \bibnamefont {McClean}},\ }\bibfield  {title} {\bibinfo {title} {On quantum backpropagation, information reuse, and cheating measurement collapse},\ }in\ \href {https://openreview.net/forum?id=HF6bnhfSqH} {\emph {\bibinfo {booktitle} {37th Conf. on Neural Information Processing Systems}}}\ (\bibinfo {year} {2023})\BibitemShut {NoStop}%
\bibitem [{\citenamefont {Gilyén}\ \emph {et~al.}(2019{\natexlab{a}})\citenamefont {Gilyén}, \citenamefont {Arunachalam},\ and\ \citenamefont {Wiebe}}]{gilyen2017OptQOptAlgGrad}%
  \BibitemOpen
  \bibfield  {author} {\bibinfo {author} {\bibfnamefont {A.}~\bibnamefont {Gilyén}}, \bibinfo {author} {\bibfnamefont {S.}~\bibnamefont {Arunachalam}},\ and\ \bibinfo {author} {\bibfnamefont {N.}~\bibnamefont {Wiebe}},\ }\bibfield  {title} {\bibinfo {title} {Optimizing quantum optimization algorithms via faster quantum gradient computation},\ }in\ \href@noop {} {\emph {\bibinfo {booktitle} {\soda{30th}}}}\ (\bibinfo {year} {2019})\ pp.\ \bibinfo {pages} {1425--1444},\ \bibinfo {note} {arxiv:1711.00465}\BibitemShut {NoStop}%
\bibitem [{\citenamefont {Sayginel}\ \emph {et~al.}(2023)\citenamefont {Sayginel}, \citenamefont {Jamet}, \citenamefont {Agarwal}, \citenamefont {Browne},\ and\ \citenamefont {Rungger}}]{sayginel2024faulttolerant}%
  \BibitemOpen
  \bibfield  {author} {\bibinfo {author} {\bibfnamefont {H.}~\bibnamefont {Sayginel}}, \bibinfo {author} {\bibfnamefont {F.}~\bibnamefont {Jamet}}, \bibinfo {author} {\bibfnamefont {A.}~\bibnamefont {Agarwal}}, \bibinfo {author} {\bibfnamefont {D.~E.}\ \bibnamefont {Browne}},\ and\ \bibinfo {author} {\bibfnamefont {I.}~\bibnamefont {Rungger}},\ }\bibfield  {title} {\bibinfo {title} {A fault-tolerant variational quantum algorithm with limited {T}-depth},\ }\href {https://doi.org/10.1088/2058-9565/ad0571} {\bibfield  {journal} {\bibinfo  {journal} {Quantum Sci. Technol.}\ }\textbf {\bibinfo {volume} {9}},\ \bibinfo {pages} {015015} (\bibinfo {year} {2023})}\BibitemShut {NoStop}%
\bibitem [{\citenamefont {Grinko}\ \emph {et~al.}(2021)\citenamefont {Grinko}, \citenamefont {Gacon}, \citenamefont {Zoufal},\ and\ \citenamefont {Woerner}}]{grinko2021iterative}%
  \BibitemOpen
  \bibfield  {author} {\bibinfo {author} {\bibfnamefont {D.}~\bibnamefont {Grinko}}, \bibinfo {author} {\bibfnamefont {J.}~\bibnamefont {Gacon}}, \bibinfo {author} {\bibfnamefont {C.}~\bibnamefont {Zoufal}},\ and\ \bibinfo {author} {\bibfnamefont {S.}~\bibnamefont {Woerner}},\ }\bibfield  {title} {\bibinfo {title} {Iterative quantum amplitude estimation},\ }\href {https://doi.org/10.1038/s41534-021-00379-1} {\bibfield  {journal} {\bibinfo  {journal} {npj Quantum Inf.}\ }\textbf {\bibinfo {volume} {7}},\ \bibinfo {pages} {52} (\bibinfo {year} {2021})}\BibitemShut {NoStop}%
\bibitem [{\citenamefont {Jnane}\ \emph {et~al.}(2024)\citenamefont {Jnane}, \citenamefont {Steinberg}, \citenamefont {Cai}, \citenamefont {Nguyen},\ and\ \citenamefont {Koczor}}]{PRXQuantum.5.010324}%
  \BibitemOpen
  \bibfield  {author} {\bibinfo {author} {\bibfnamefont {H.}~\bibnamefont {Jnane}}, \bibinfo {author} {\bibfnamefont {J.}~\bibnamefont {Steinberg}}, \bibinfo {author} {\bibfnamefont {Z.}~\bibnamefont {Cai}}, \bibinfo {author} {\bibfnamefont {H.~C.}\ \bibnamefont {Nguyen}},\ and\ \bibinfo {author} {\bibfnamefont {B.}~\bibnamefont {Koczor}},\ }\bibfield  {title} {\bibinfo {title} {Quantum error mitigated classical shadows},\ }\href {https://doi.org/10.1103/PRXQuantum.5.010324} {\bibfield  {journal} {\bibinfo  {journal} {PRX Quantum}\ }\textbf {\bibinfo {volume} {5}},\ \bibinfo {pages} {010324} (\bibinfo {year} {2024})}\BibitemShut {NoStop}%
\bibitem [{\citenamefont {Huang}\ \emph {et~al.}(2020)\citenamefont {Huang}, \citenamefont {Kueng},\ and\ \citenamefont {Preskill}}]{Huang_2020}%
  \BibitemOpen
  \bibfield  {author} {\bibinfo {author} {\bibfnamefont {H.-Y.}\ \bibnamefont {Huang}}, \bibinfo {author} {\bibfnamefont {R.}~\bibnamefont {Kueng}},\ and\ \bibinfo {author} {\bibfnamefont {J.}~\bibnamefont {Preskill}},\ }\bibfield  {title} {\bibinfo {title} {Predicting many properties of a quantum system from very few measurements},\ }\href {https://doi.org/10.1038/s41567-020-0932-7} {\bibfield  {journal} {\bibinfo  {journal} {Nat. Phys.}\ }\textbf {\bibinfo {volume} {16}},\ \bibinfo {pages} {1050–1057} (\bibinfo {year} {2020})}\BibitemShut {NoStop}%
\bibitem [{\citenamefont {Boyd}\ and\ \citenamefont {Koczor}(2022)}]{PhysRevX.12.041022}%
  \BibitemOpen
  \bibfield  {author} {\bibinfo {author} {\bibfnamefont {G.}~\bibnamefont {Boyd}}\ and\ \bibinfo {author} {\bibfnamefont {B.}~\bibnamefont {Koczor}},\ }\bibfield  {title} {\bibinfo {title} {Training variational quantum circuits with {C}o{V}a{R}: Covariance root finding with classical shadows},\ }\href {https://doi.org/10.1103/PhysRevX.12.041022} {\bibfield  {journal} {\bibinfo  {journal} {Phys. Rev. X}\ }\textbf {\bibinfo {volume} {12}},\ \bibinfo {pages} {041022} (\bibinfo {year} {2022})}\BibitemShut {NoStop}%
\bibitem [{\citenamefont {Shor}(1997)}]{shor1994Factoring}%
  \BibitemOpen
  \bibfield  {author} {\bibinfo {author} {\bibfnamefont {P.~W.}\ \bibnamefont {Shor}},\ }\bibfield  {title} {\bibinfo {title} {Polynomial-time algorithms for prime factorization and discrete logarithms on a quantum computer},\ }\href {https://doi.org/10.1137/S0097539795293172} {\bibfield  {journal} {\bibinfo  {journal} {\siamjc}\ }\textbf {\bibinfo {volume} {26}},\ \bibinfo {pages} {1484} (\bibinfo {year} {1997})},\ \bibinfo {note} {earlier version in FOCS'94., arXiv:quant-ph/9508027}\BibitemShut {NoStop}%
\bibitem [{\citenamefont {Kitaev}(1995)}]{kitaev1995QMeasAndAbelianStabilizer}%
  \BibitemOpen
  \bibfield  {author} {\bibinfo {author} {\bibfnamefont {A.~Y.}\ \bibnamefont {Kitaev}},\ }\href@noop {} {\bibinfo {title} {Quantum measurements and the {A}belian stabilizer problem}} (\bibinfo {year} {1995}),\ \Eprint {https://arxiv.org/abs/quant-ph/9511026} {arXiv:quant-ph/9511026 [quant-ph]} \BibitemShut {NoStop}%
\bibitem [{\citenamefont {Dalzell}\ \emph {et~al.}(2024)\citenamefont {Dalzell}, \citenamefont {McArdle}, \citenamefont {Berta}, \citenamefont {Bienias}, \citenamefont {Chen}, \citenamefont {Gilyén}, \citenamefont {Hann}, \citenamefont {Kastoryano}, \citenamefont {Khabiboulline}, \citenamefont {Kubica}, \citenamefont {Salton}, \citenamefont {Wang},\ and\ \citenamefont {Brandão}}]{dalzell2023QuantumAlgSurvey}%
  \BibitemOpen
  \bibfield  {author} {\bibinfo {author} {\bibfnamefont {A.~M.}\ \bibnamefont {Dalzell}}, \bibinfo {author} {\bibfnamefont {S.}~\bibnamefont {McArdle}}, \bibinfo {author} {\bibfnamefont {M.}~\bibnamefont {Berta}}, \bibinfo {author} {\bibfnamefont {P.}~\bibnamefont {Bienias}}, \bibinfo {author} {\bibfnamefont {C.-F.}\ \bibnamefont {Chen}}, \bibinfo {author} {\bibfnamefont {A.}~\bibnamefont {Gilyén}}, \bibinfo {author} {\bibfnamefont {C.~T.}\ \bibnamefont {Hann}}, \bibinfo {author} {\bibfnamefont {M.~J.}\ \bibnamefont {Kastoryano}}, \bibinfo {author} {\bibfnamefont {E.~T.}\ \bibnamefont {Khabiboulline}}, \bibinfo {author} {\bibfnamefont {A.}~\bibnamefont {Kubica}}, \bibinfo {author} {\bibfnamefont {G.}~\bibnamefont {Salton}}, \bibinfo {author} {\bibfnamefont {S.}~\bibnamefont {Wang}},\ and\ \bibinfo {author} {\bibfnamefont {F.~G. S.~L.}\ \bibnamefont {Brandão}},\ }\href@noop {} {\emph {\bibinfo {title} {Quantum algorithms: A survey of applications and end-to-end complexities}}}\ (\bibinfo  {publisher}
  {Cambridge University Press},\ \bibinfo {year} {2024})\ \bibinfo {note} {in print, forthcoming, arXiv:2310.03011}\BibitemShut {NoStop}%
\bibitem [{\citenamefont {Lee}\ \emph {et~al.}(2023)\citenamefont {Lee}, \citenamefont {Lee}, \citenamefont {Zhai}, \citenamefont {Tong}, \citenamefont {Dalzell}, \citenamefont {Kumar}, \citenamefont {Helms}, \citenamefont {Gray}, \citenamefont {Cui}, \citenamefont {Liu}, \citenamefont {Kastoryano}, \citenamefont {Babbush}, \citenamefont {Preskill}, \citenamefont {Reichman}, \citenamefont {Campbell}, \citenamefont {Valeev}, \citenamefont {Lin},\ and\ \citenamefont {Chan}}]{lee2023EvidenceQuantAdvGroundState}%
  \BibitemOpen
  \bibfield  {author} {\bibinfo {author} {\bibfnamefont {S.}~\bibnamefont {Lee}}, \bibinfo {author} {\bibfnamefont {J.}~\bibnamefont {Lee}}, \bibinfo {author} {\bibfnamefont {H.}~\bibnamefont {Zhai}}, \bibinfo {author} {\bibfnamefont {Y.}~\bibnamefont {Tong}}, \bibinfo {author} {\bibfnamefont {A.~M.}\ \bibnamefont {Dalzell}}, \bibinfo {author} {\bibfnamefont {A.}~\bibnamefont {Kumar}}, \bibinfo {author} {\bibfnamefont {P.}~\bibnamefont {Helms}}, \bibinfo {author} {\bibfnamefont {J.}~\bibnamefont {Gray}}, \bibinfo {author} {\bibfnamefont {Z.-H.}\ \bibnamefont {Cui}}, \bibinfo {author} {\bibfnamefont {W.}~\bibnamefont {Liu}}, \bibinfo {author} {\bibfnamefont {M.}~\bibnamefont {Kastoryano}}, \bibinfo {author} {\bibfnamefont {R.}~\bibnamefont {Babbush}}, \bibinfo {author} {\bibfnamefont {J.}~\bibnamefont {Preskill}}, \bibinfo {author} {\bibfnamefont {D.~R.}\ \bibnamefont {Reichman}}, \bibinfo {author} {\bibfnamefont {E.~T.}\ \bibnamefont {Campbell}}, \bibinfo {author} {\bibfnamefont {E.~F.}\ \bibnamefont
  {Valeev}}, \bibinfo {author} {\bibfnamefont {L.}~\bibnamefont {Lin}},\ and\ \bibinfo {author} {\bibfnamefont {G.~K.-L.}\ \bibnamefont {Chan}},\ }\bibfield  {title} {\bibinfo {title} {Evaluating the evidence for exponential quantum advantage in ground-state quantum chemistry},\ }\href {https://doi.org/10.1038/s41467-023-37587-6} {\bibfield  {journal} {\bibinfo  {journal} {Nat. Commun.}\ }\textbf {\bibinfo {volume} {14}},\ \bibinfo {pages} {1952} (\bibinfo {year} {2023})}\BibitemShut {NoStop}%
\bibitem [{\citenamefont {Childs}\ \emph {et~al.}(2018)\citenamefont {Childs}, \citenamefont {Maslov}, \citenamefont {Nam}, \citenamefont {Ross},\ and\ \citenamefont {Su}}]{childs2018toward}%
  \BibitemOpen
  \bibfield  {author} {\bibinfo {author} {\bibfnamefont {A.~M.}\ \bibnamefont {Childs}}, \bibinfo {author} {\bibfnamefont {D.}~\bibnamefont {Maslov}}, \bibinfo {author} {\bibfnamefont {Y.}~\bibnamefont {Nam}}, \bibinfo {author} {\bibfnamefont {N.~J.}\ \bibnamefont {Ross}},\ and\ \bibinfo {author} {\bibfnamefont {Y.}~\bibnamefont {Su}},\ }\bibfield  {title} {\bibinfo {title} {Toward the first quantum simulation with quantum speedup},\ }\href {https://doi.org/10.1073/pnas.1801723115} {\bibfield  {journal} {\bibinfo  {journal} {Proc. Natl. Acad. Sci. U.S.A.}\ }\textbf {\bibinfo {volume} {115}},\ \bibinfo {pages} {9456} (\bibinfo {year} {2018})}\BibitemShut {NoStop}%
\bibitem [{\citenamefont {Childs}\ \emph {et~al.}(2021)\citenamefont {Childs}, \citenamefont {Su}, \citenamefont {Tran}, \citenamefont {Wiebe},\ and\ \citenamefont {Zhu}}]{childs2021theory}%
  \BibitemOpen
  \bibfield  {author} {\bibinfo {author} {\bibfnamefont {A.~M.}\ \bibnamefont {Childs}}, \bibinfo {author} {\bibfnamefont {Y.}~\bibnamefont {Su}}, \bibinfo {author} {\bibfnamefont {M.~C.}\ \bibnamefont {Tran}}, \bibinfo {author} {\bibfnamefont {N.}~\bibnamefont {Wiebe}},\ and\ \bibinfo {author} {\bibfnamefont {S.}~\bibnamefont {Zhu}},\ }\bibfield  {title} {\bibinfo {title} {Theory of {T}rotter error with commutator scaling},\ }\href {https://doi.org/10.1103/PhysRevX.11.011020} {\bibfield  {journal} {\bibinfo  {journal} {Phys. Rev. X}\ }\textbf {\bibinfo {volume} {11}},\ \bibinfo {pages} {011020} (\bibinfo {year} {2021})}\BibitemShut {NoStop}%
\bibitem [{\citenamefont {Rubin}\ \emph {et~al.}(2024)\citenamefont {Rubin}, \citenamefont {Berry}, \citenamefont {Kononov}, \citenamefont {Malone}, \citenamefont {Khattar}, \citenamefont {White}, \citenamefont {Lee}, \citenamefont {Neven}, \citenamefont {Babbush},\ and\ \citenamefont {Baczewski}}]{rubin2024quantum}%
  \BibitemOpen
  \bibfield  {author} {\bibinfo {author} {\bibfnamefont {N.~C.}\ \bibnamefont {Rubin}}, \bibinfo {author} {\bibfnamefont {D.~W.}\ \bibnamefont {Berry}}, \bibinfo {author} {\bibfnamefont {A.}~\bibnamefont {Kononov}}, \bibinfo {author} {\bibfnamefont {F.~D.}\ \bibnamefont {Malone}}, \bibinfo {author} {\bibfnamefont {T.}~\bibnamefont {Khattar}}, \bibinfo {author} {\bibfnamefont {A.}~\bibnamefont {White}}, \bibinfo {author} {\bibfnamefont {J.}~\bibnamefont {Lee}}, \bibinfo {author} {\bibfnamefont {H.}~\bibnamefont {Neven}}, \bibinfo {author} {\bibfnamefont {R.}~\bibnamefont {Babbush}},\ and\ \bibinfo {author} {\bibfnamefont {A.~D.}\ \bibnamefont {Baczewski}},\ }\bibfield  {title} {\bibinfo {title} {Quantum computation of stopping power for inertial fusion target design},\ }\href {https://doi.org/10.1073/pnas.2317772121} {\bibfield  {journal} {\bibinfo  {journal} {Proc. Natl. Acad. Sci. U.S.A.}\ }\textbf {\bibinfo {volume} {121}},\ \bibinfo {pages} {e2317772121} (\bibinfo {year} {2024})}\BibitemShut {NoStop}%
\bibitem [{\citenamefont {Rosenberg}\ \emph {et~al.}(2024)\citenamefont {Rosenberg} \emph {et~al.}}]{Rosenberg_2024}%
  \BibitemOpen
  \bibfield  {author} {\bibinfo {author} {\bibfnamefont {E.}~\bibnamefont {Rosenberg}} \emph {et~al.},\ }\bibfield  {title} {\bibinfo {title} {Dynamics of magnetization at infinite temperature in a {H}eisenberg spin chain},\ }\href {https://doi.org/10.1126/science.adi7877} {\bibfield  {journal} {\bibinfo  {journal} {Science}\ }\textbf {\bibinfo {volume} {384}},\ \bibinfo {pages} {48–53} (\bibinfo {year} {2024})}\BibitemShut {NoStop}%
\bibitem [{\citenamefont {Kim}\ \emph {et~al.}(2023{\natexlab{b}})\citenamefont {Kim}, \citenamefont {Eddins}, \citenamefont {Anand}, \citenamefont {Wei}, \citenamefont {Van Den~Berg}, \citenamefont {Rosenblatt}, \citenamefont {Nayfeh}, \citenamefont {Wu}, \citenamefont {Zaletel}, \citenamefont {Temme},\ and\ \citenamefont {Kandala}}]{Kim_2023}%
  \BibitemOpen
  \bibfield  {author} {\bibinfo {author} {\bibfnamefont {Y.}~\bibnamefont {Kim}}, \bibinfo {author} {\bibfnamefont {A.}~\bibnamefont {Eddins}}, \bibinfo {author} {\bibfnamefont {S.}~\bibnamefont {Anand}}, \bibinfo {author} {\bibfnamefont {K.~X.}\ \bibnamefont {Wei}}, \bibinfo {author} {\bibfnamefont {E.}~\bibnamefont {Van Den~Berg}}, \bibinfo {author} {\bibfnamefont {S.}~\bibnamefont {Rosenblatt}}, \bibinfo {author} {\bibfnamefont {H.}~\bibnamefont {Nayfeh}}, \bibinfo {author} {\bibfnamefont {Y.}~\bibnamefont {Wu}}, \bibinfo {author} {\bibfnamefont {M.}~\bibnamefont {Zaletel}}, \bibinfo {author} {\bibfnamefont {K.}~\bibnamefont {Temme}},\ and\ \bibinfo {author} {\bibfnamefont {A.}~\bibnamefont {Kandala}},\ }\bibfield  {title} {\bibinfo {title} {Evidence for the utility of quantum computing before fault tolerance},\ }\href {https://doi.org/10.1038/s41586-023-06096-3} {\bibfield  {journal} {\bibinfo  {journal} {Nature}\ }\textbf {\bibinfo {volume} {618}},\ \bibinfo {pages} {500} (\bibinfo {year}
  {2023}{\natexlab{b}})}\BibitemShut {NoStop}%
\bibitem [{\citenamefont {Fischer}\ \emph {et~al.}(2024)\citenamefont {Fischer}, \citenamefont {Leahy}, \citenamefont {Eddins}, \citenamefont {Keenan}, \citenamefont {Ferracin}, \citenamefont {Rossi}, \citenamefont {Kim}, \citenamefont {He}, \citenamefont {Pietracaprina}, \citenamefont {Sokolov}, \citenamefont {Dooley}, \citenamefont {Zimborás}, \citenamefont {Tacchino}, \citenamefont {Maniscalco}, \citenamefont {Goold}, \citenamefont {García-Pérez}, \citenamefont {Tavernelli}, \citenamefont {Kandala},\ and\ \citenamefont {Filippov}}]{Fischer_2024}%
  \BibitemOpen
  \bibfield  {author} {\bibinfo {author} {\bibfnamefont {L.~E.}\ \bibnamefont {Fischer}}, \bibinfo {author} {\bibfnamefont {M.}~\bibnamefont {Leahy}}, \bibinfo {author} {\bibfnamefont {A.}~\bibnamefont {Eddins}}, \bibinfo {author} {\bibfnamefont {N.}~\bibnamefont {Keenan}}, \bibinfo {author} {\bibfnamefont {D.}~\bibnamefont {Ferracin}}, \bibinfo {author} {\bibfnamefont {M.~A.~C.}\ \bibnamefont {Rossi}}, \bibinfo {author} {\bibfnamefont {Y.}~\bibnamefont {Kim}}, \bibinfo {author} {\bibfnamefont {A.}~\bibnamefont {He}}, \bibinfo {author} {\bibfnamefont {F.}~\bibnamefont {Pietracaprina}}, \bibinfo {author} {\bibfnamefont {B.}~\bibnamefont {Sokolov}}, \bibinfo {author} {\bibfnamefont {S.}~\bibnamefont {Dooley}}, \bibinfo {author} {\bibfnamefont {Z.}~\bibnamefont {Zimborás}}, \bibinfo {author} {\bibfnamefont {F.}~\bibnamefont {Tacchino}}, \bibinfo {author} {\bibfnamefont {S.}~\bibnamefont {Maniscalco}}, \bibinfo {author} {\bibfnamefont {J.}~\bibnamefont {Goold}}, \bibinfo {author} {\bibfnamefont {G.}~\bibnamefont
  {García-Pérez}}, \bibinfo {author} {\bibfnamefont {I.}~\bibnamefont {Tavernelli}}, \bibinfo {author} {\bibfnamefont {A.}~\bibnamefont {Kandala}},\ and\ \bibinfo {author} {\bibfnamefont {S.~N.}\ \bibnamefont {Filippov}},\ }\href@noop {} {\bibinfo {title} {Dynamical simulations of many-body quantum chaos on a quantum computer}} (\bibinfo {year} {2024}),\ \Eprint {https://arxiv.org/abs/2411.00765} {arXiv:2411.00765 [quant-ph]} \BibitemShut {NoStop}%
\bibitem [{\citenamefont {Li}\ and\ \citenamefont {Benjamin}(2017)}]{li2017VariaSim}%
  \BibitemOpen
  \bibfield  {author} {\bibinfo {author} {\bibfnamefont {Y.}~\bibnamefont {Li}}\ and\ \bibinfo {author} {\bibfnamefont {S.~C.}\ \bibnamefont {Benjamin}},\ }\bibfield  {title} {\bibinfo {title} {Efficient variational quantum simulator incorporating active error minimization},\ }\href {https://doi.org/10.1103/PhysRevX.7.021050} {\bibfield  {journal} {\bibinfo  {journal} {Phys. Rev. X}\ }\textbf {\bibinfo {volume} {7}},\ \bibinfo {pages} {021050} (\bibinfo {year} {2017})}\BibitemShut {NoStop}%
\bibitem [{\citenamefont {Bosse}\ \emph {et~al.}(2024)\citenamefont {Bosse}, \citenamefont {Childs}, \citenamefont {Derby}, \citenamefont {Gambetta}, \citenamefont {Montanaro},\ and\ \citenamefont {Santos}}]{Bosse24}%
  \BibitemOpen
  \bibfield  {author} {\bibinfo {author} {\bibfnamefont {J.~L.}\ \bibnamefont {Bosse}}, \bibinfo {author} {\bibfnamefont {A.~M.}\ \bibnamefont {Childs}}, \bibinfo {author} {\bibfnamefont {C.}~\bibnamefont {Derby}}, \bibinfo {author} {\bibfnamefont {F.~M.}\ \bibnamefont {Gambetta}}, \bibinfo {author} {\bibfnamefont {A.}~\bibnamefont {Montanaro}},\ and\ \bibinfo {author} {\bibfnamefont {R.~A.}\ \bibnamefont {Santos}},\ }\href@noop {} {\bibinfo {title} {Efficient and practical {H}amiltonian simulation from time-dependent product formulas}} (\bibinfo {year} {2024}),\ \Eprint {https://arxiv.org/abs/2403.08729} {arXiv:2403.08729 [quant-ph]} \BibitemShut {NoStop}%
\bibitem [{\citenamefont {Low}\ and\ \citenamefont {Chuang}(2017)}]{low2016HamSimQSignProc}%
  \BibitemOpen
  \bibfield  {author} {\bibinfo {author} {\bibfnamefont {G.~H.}\ \bibnamefont {Low}}\ and\ \bibinfo {author} {\bibfnamefont {I.~L.}\ \bibnamefont {Chuang}},\ }\bibfield  {title} {\bibinfo {title} {Optimal {H}amiltonian simulation by quantum signal processing},\ }\href {https://doi.org/10.1103/PhysRevLett.118.010501} {\bibfield  {journal} {\bibinfo  {journal} {\prl}\ }\textbf {\bibinfo {volume} {118}},\ \bibinfo {pages} {010501} (\bibinfo {year} {2017})}\BibitemShut {NoStop}%
\bibitem [{\citenamefont {Gilyén}\ \emph {et~al.}(2019{\natexlab{b}})\citenamefont {Gilyén}, \citenamefont {Su}, \citenamefont {Low},\ and\ \citenamefont {Wiebe}}]{gilyen2018QSingValTransf}%
  \BibitemOpen
  \bibfield  {author} {\bibinfo {author} {\bibfnamefont {A.}~\bibnamefont {Gilyén}}, \bibinfo {author} {\bibfnamefont {Y.}~\bibnamefont {Su}}, \bibinfo {author} {\bibfnamefont {G.~H.}\ \bibnamefont {Low}},\ and\ \bibinfo {author} {\bibfnamefont {N.}~\bibnamefont {Wiebe}},\ }\bibfield  {title} {\bibinfo {title} {Quantum singular value transformation and beyond: {E}xponential improvements for quantum matrix arithmetics},\ }in\ \href@noop {} {\emph {\bibinfo {booktitle} {\stoc{51st}}}}\ (\bibinfo {year} {2019})\ pp.\ \bibinfo {pages} {193--204},\ \bibinfo {note} {full version in arXiv:1806.01838}\BibitemShut {NoStop}%
\bibitem [{\citenamefont {Dong}\ \emph {et~al.}(2022)\citenamefont {Dong}, \citenamefont {Lin},\ and\ \citenamefont {Tong}}]{PRXQuantum.3.040305}%
  \BibitemOpen
  \bibfield  {author} {\bibinfo {author} {\bibfnamefont {Y.}~\bibnamefont {Dong}}, \bibinfo {author} {\bibfnamefont {L.}~\bibnamefont {Lin}},\ and\ \bibinfo {author} {\bibfnamefont {Y.}~\bibnamefont {Tong}},\ }\bibfield  {title} {\bibinfo {title} {Ground-state preparation and energy estimation on early fault-tolerant quantum computers via quantum eigenvalue transformation of unitary matrices},\ }\href {https://doi.org/10.1103/PRXQuantum.3.040305} {\bibfield  {journal} {\bibinfo  {journal} {PRX Quantum}\ }\textbf {\bibinfo {volume} {3}},\ \bibinfo {pages} {040305} (\bibinfo {year} {2022})}\BibitemShut {NoStop}%
\bibitem [{\citenamefont {Chen}\ \emph {et~al.}(2023)\citenamefont {Chen}, \citenamefont {Kastoryano}, \citenamefont {Brandão},\ and\ \citenamefont {Gilyén}}]{chen2023QThermalStatePrep}%
  \BibitemOpen
  \bibfield  {author} {\bibinfo {author} {\bibfnamefont {C.-F.}\ \bibnamefont {Chen}}, \bibinfo {author} {\bibfnamefont {M.~J.}\ \bibnamefont {Kastoryano}}, \bibinfo {author} {\bibfnamefont {F.~G. S.~L.}\ \bibnamefont {Brandão}},\ and\ \bibinfo {author} {\bibfnamefont {A.}~\bibnamefont {Gilyén}},\ }\href@noop {} {\bibinfo {title} {Quantum thermal state preparation}} (\bibinfo {year} {2023}),\ \Eprint {https://arxiv.org/abs/2303.18224} {arXiv:2303.18224 [quant-ph]} \BibitemShut {NoStop}%
\bibitem [{\citenamefont {Cubitt}(2023)}]{cubitt2023dissipativegroundstatepreparation}%
  \BibitemOpen
  \bibfield  {author} {\bibinfo {author} {\bibfnamefont {T.~S.}\ \bibnamefont {Cubitt}},\ }\href@noop {} {\bibinfo {title} {Dissipative ground state preparation and the {D}issipative {Q}uantum {E}igensolver}} (\bibinfo {year} {2023}),\ \Eprint {https://arxiv.org/abs/2303.11962} {arXiv:2303.11962 [quant-ph]} \BibitemShut {NoStop}%
\bibitem [{\citenamefont {Allen}\ \emph {et~al.}(2024)\citenamefont {Allen}, \citenamefont {Machado}, \citenamefont {Chuang}, \citenamefont {Huang},\ and\ \citenamefont {Choi}}]{allen2024sensing}%
  \BibitemOpen
  \bibfield  {author} {\bibinfo {author} {\bibfnamefont {R.~R.}\ \bibnamefont {Allen}}, \bibinfo {author} {\bibfnamefont {F.}~\bibnamefont {Machado}}, \bibinfo {author} {\bibfnamefont {I.~L.}\ \bibnamefont {Chuang}}, \bibinfo {author} {\bibfnamefont {H.-Y.}\ \bibnamefont {Huang}},\ and\ \bibinfo {author} {\bibfnamefont {S.}~\bibnamefont {Choi}},\ }\bibfield  {title} {\bibinfo {title} {Quantum computing enhanced sensing}} (\bibinfo {year} {2024}),\ \bibinfo {note} {unpublished}\BibitemShut {NoStop}%
\end{thebibliography}%

\clearpage
\appendix

\end{document}